\documentclass[bibyear]{aa}  
\usepackage[justification=centering]{caption}
\usepackage[rightcaption]{sidecap}
\usepackage{subcaption}
\usepackage{graphicx}
\usepackage{xcolor}
\usepackage{natbib}
\usepackage{color}
\usepackage{threeparttable}

\usepackage[singlelinecheck=false,justification=justified]{caption}
\usepackage{txfonts}

\usepackage{gensymb}

\usepackage{hyperref}
\hypersetup{
    colorlinks=true, 
    linkcolor=blue, 
    citecolor=blue, 
    urlcolor=blue 
}
\usepackage{mathtools}
\usepackage{placeins}
\usepackage{bm}
\usepackage{esvect}
\begin{document}

    \title{PLATOSpec's first results: planets WASP-35b and TOI-622b are on aligned orbits, and K2-237b is on a polar orbit}
    \titlerunning{PLATOSpec finds a planet on a polar orbit}
    
   \author{J. Zak
          \inst{1}
            \and
           P. Kabath \inst{1}
           \and
        H.\,M.\,J. Boffin \inst{2}
                       \and
                       J. Liptak \inst{1, 3}
                       \and 
                       M. Skarka \inst{1} \and R. Brahm \inst{4, 5}
                       \and P. Gajdo{\v s} \inst{1, 6} \and
           A. Bocchieri \inst{7}   \and D. Itrich \inst{8} \and L.~Vanzi \inst{9, 5}\and P. Pintr \inst{10} \and J. Janik \inst{11} \and A. Hatzes \inst{12}
           }

\institute{
    Astronomical Institute of the Czech Academy of Sciences, Fri\v{c}ova 298, 25165 Ond\v{r}ejov, Czech Republic;
    \email{zak@asu.cas.cz}
    \and
    European Southern Observatory, Karl-Schwarzschild-str. 2, 85748 Garching, Germany
    \and
    Astronomical Institute of Charles University, V Hole\v{s}ovi\v{c}k\'{a}ch 2, CZ-180 00 Prague, Czech Republic
    \and
School of engineering and sciences, Universidad Adolfo Ibañez, Av. Diag. Las Torres 2640, 7941169 Santiago, Chile
\and
Center of Astro Engineering, Pontificia Universidad Católica de Chile, Av. Vicuña Mackenna 4860, 782-043 Santiago, Chile
\and
Institute of Physics, Faculty of Science, Pavol Jozef Šafárik University, Park Angelinum 9, 040 01 Košice, Slovakia
\and
    Dipartimento di Fisica, La Sapienza Università di Roma, Piazzale Aldo Moro 5, Roma, 00185, Italy
\and
           Steward Observatory, The University of Arizona, 933 N. Cherry Ave, Tucson, AZ 85721, USA
\and
           Department of Electrical Engineering, Pontificia Universidad Católica de Chile, Av. Vicuña Mackenna 4860, 782-043 Santiago, Chile \and
           Department of Theoretical Physics and Astrophysics, Masaryk Univesity, Kotl\'a\v{r}sk\'a 2, 60200 Brno, Czech Republic
\and
           Institute of Plasma Physics of Czech Academy of Sciences, U Slovanky 2525/1a, 182 00 Praha 8, Czech Republic
\and
Thüringer Landessternwarte, D-07778 Tautenburg, Germany
}

   \date{Received June 28, 2025; accepted August XX, 2025}


  \abstract
   {The spin-orbit angle between a stellar spin axis and its planetary orbital axis is a key diagnostic of planetary migration pathways, yet the mechanisms shaping the observed spin-orbit distribution remain incompletely understood. Combining the spin-orbit angle with atmospheric measurements has emerged as a powerful method of studying exoplanets that showcases the synergy between ground- and space-based observations. We present the Rossiter-McLaughlin effect measurements of the projected spin-orbit angle ($\lambda$) for three gaseous exoplanets using the newly commissioned PLATOSpec instrument on the E152 Telescope at La Silla Observatory. For WASP-35b, we determine $\lambda = 1_{-18}^{+19}$~deg, demonstrating PLATOSpec's capabilities through excellent agreement with HARPS-N literature data. We provide the first spin-orbit measurements for TOI-622b ($\lambda =-4 \pm 12$\,deg, true spin-orbit angle $\psi = $16.1$^{+8.0}_{-9.7}$\,deg), revealing an aligned orbit consistent with quiescent disc migration. For K2-237b, we find $\lambda = 91 \pm 7$~deg and $\psi = $90.5$^{+6.8}_{-6.2}$\,deg, indicating a nearly perfect polar orbit, which suggests a history consistent with disc-free migration, contrasting previous studies inferring disc migration. TOI-622b populates a sparsely populated region of sub-Jovian planets with measured spin-orbit angles orbiting stars above the Kraft break, while K2-237b's polar configuration strengthens tentative evidence for preferential orbital orientations. All three systems are compelling targets for future atmospheric characterization, where these dynamical constraints will be vital for a comprehensive understanding of their formation and evolution.}

   \keywords{Techniques: radial velocities  --
               Planets and satellites: gaseous planets -- Planets and satellites: atmospheres--
                Planet-star interactions -- Planets and satellites: individual: WASP-35b, TOI-622b, K2-237b
               }

   \maketitle
%

\section{Introduction} \label{sec:intro}
The upcoming PLAnetary Transits and Oscillations of stars ({\it PLATO}) mission \citep{rau25} is expected to discover tens of thousands of exoplanetary candidates. To facilitate their vetting and characterization, dedicated instruments and projects are being developed, such as the Ground-based Observation Programme\footnote{\url{https://warwick.ac.uk/fac/sci/physics/research/astro/plato-science/research/researchareas/followup/}}. PLATOSpec\footnote{\url{https://stel.asu.cas.cz/plato/spectrograph.html}} (Kabath et al., in prep.) is a new high-resolution spectrograph that was developed and commissioned in order to facilitate ground-based follow-up of space missions. Besides performing stellar characterization and measuring the planetary masses and eccentricities, PLATOSpec will be able to provide the projected angles between the orbital plane and the rotation axis of the star (or spin-orbit angles) -- a key aspect of the orbital architecture \citep{alb22}.

To explain the observed population of exoplanets on short orbits, an unknown fraction of gaseous planets  is expected to undertake migration after their formation beyond the snow line towards the host star  \citep{ida04,fab07,baru14}. The two main migration scenarios are disc-free migration (high-eccentricity migration) and disc migration \citep{baru14,daw18}. However, the occurrence rate of each migration mechanism is still poorly understood. This precludes us from refining our planet formation models in great detail and describing the observed demographics of exoplanets \citep{madhu16,ray22}.

The spin-orbit angle has emerged in recent years as a powerful probe to study exoplanets and their evolution \citep{tri17}. Complementarily to other physical and orbital parameters, it provides information about the system's history. The Solar System has a low spin-orbit angle of around 7 degrees \citep{bec05}. This non-zero value is usually attributed to the non-quiet history of the Solar System \citep{bai16}. The projected spin-orbit angle ($\lambda$) is most commonly measured using the Rossiter-McLaughlin (R-M) effect \citep{ross24,mcl} with high-resolution spectroscopy. Previous studies have revealed various orientations of the planetary orbits, ranging from perfectly aligned to orbits with polar and retrograde orientation suggesting numerous mechanisms are sculpting these orbits \citep{cri14}. The availability of precise high-resolution data has led to an increase in the available measurements in recent years \citep[e.g.,][]{bou23, zak24a}.
The number of projected spin-orbit angle measurements approaches 300\footnote{Retrieved from the TEPCat catalog \citep{south11}.}. Despite this abundance, the understanding of the distribution of these angles remains elusive as several trends were proposed but not confirmed. 


The distribution of spin-orbit angles appears to be shaped by distinct physical processes depending on stellar and planetary properties. For hot Jupiters, a clear trend has been observed: those orbiting cool stars tend to have their orbits aligned, while those around hot stars exhibit a broad range of spin-orbit angles \citep{winn10}. This pattern is widely thought to be a consequence of tidal realignment, a mechanism that is less effective for planets orbiting hot stars with radiative envelopes \citep{alb12}. However, our understanding is limited when it comes to sub-Jovian planets, as there is a notable lack of spin-orbit measurements for this population around hot stars, making a direct comparison impossible. Framework proposed by \citet{hix23} suggests that hotter, more massive stars are more likely to form multiple planets. Subsequent planet-planet scattering and secular interactions within these systems could then excite large spin-orbit misalignment. This hypothesis leads to the prediction that a significant portion of low-mass planets orbiting hot stars should have misaligned orbits.

\citet{alb21} suggested that certain orbital architectures might be preferred, especially around the polar configuration.
Several mechanisms \citep[e.g.,][]{lai12,chen24,lo24} were suggested to explain this trend, but more measurements are needed to verify such inferences as well as to break degeneracies and deliver a more complete understanding of, e.g., the origin of aligned but eccentric Warm Jupiters \citep{rice22wj,esp23,bie25}. It is highly warranted to obtain measurements of misaligned planets covering the full parameter space to avoid biases \citep{esp24nep} and check for differences between various exoplanetary populations to address whether the mechanisms shaping the spin-orbit distribution of Jovian planets are the same as for sub-Jovian planets. For example, \citet{pet20} suggested disc-driven resonance mechanism exciting the orbit of Neptunian planets with an outer gas giant. Such a mechanism is expected to be inefficient for Jovian planets.


In this work, we present three projected spin-orbit angle measurements of gaseous planets with the newly commissioned PLATOSpec spectrograph. All three are on the Tier 3 (planets selected for detailed studies) target candidate list of the \textit{Ariel} mission and represent good targets for atmospheric characterization.

\section{Data sets and their analyses} \label{sec:data}

\subsection{Datasets}
We have used the PLATOSpec instrument (Kabath et al., in prep.) mounted on the ESO 1.52 m telescope (E152) at the La Silla Observatory, Chile\footnote{\url{https://www.eso.org/public/teles-instr/lasilla/152metre/}}. PLATOSpec is a fiber-fed echelle spectrograph placed in a thermally controlled environment that uses simultaneous ThAr wavelength calibration to provide precise RV measurements. As presented in \citet{kab22ps}, the main role of PLATOSpec will be to provide support to space-based missions such as \textit{PLATO} and \textit{Ariel}. Among the primary science goals will be planetary candidate vetting and detection and subsequent characterization of gas giant planets. This is nicely illustrated in our target selection as one of the systems (TOI-622) is located in the LOPS2\footnote{Long-duration Observation Phase} \textit{PLATO} field \citep{nasc25}. The wavelength coverage of the instrument spans from 380 to 680 nm, with a resolving power of $R \approx 70\,000$, corresponding to 4.3\,km\,s$^{-1}$ per resolution element. Table \ref{obs_logs} shows the properties of the data sets we used. We combined the results from multiple nights when available; no RV offsets between the nights were included in the fit. The PLATOSpec spectra were reduced by the CERES+ pipeline based on the CERES pipeline \citep{brahm17}. The spectra are already corrected to the Solar system barycentric frame of reference. The radial velocities were also obtained from the CERES+ pipeline together with their uncertainties and are listed in Appendix \ref{appB} in Tables \ref{table:rv35} to \ref{table:rvk2237}. Table \ref{tab:targets_compiled} displays the properties of the systems we study.

\begin{table*}[htbp]
\captionsetup{justification=centering}
\caption{Observing logs for the three systems. The number in parenthesis represents the number of frames taken in-transit.}
\vspace{-.4cm}
\label{obs_logs}
\centering 
\begin{center}
\begin{tabular}{lccccc}
\hline
Target & Night & No.     & Exp.     & Airmass & Median  \\ 
     & Obs~&~frames~& Time (s) & range    &  SNR$^a$    \\ 
\hline
WASP-35 & 2025-01-05/06~&~17 (10)~& 1200     & 1.17-1.09-1.97~&~28-50  \\ 
 & 2025-01-24/25~&~17 (11)~& 900-1020     & 1.09-2.20~&~22-33 \\

TOI-622 & 2025-02-09/10~&~11 (10)~& 900     & 1.17-1.84~&~96-110 \\ 

& 2025-04-27/28~&~17 (14)~& 900     & 1.07-1.86~&~90-141 \\ 

K2-237 & 2025-03-17/18~&~10 (6)~& 1200-1750     &1.67-1.00~&~23-52\\

 & 2025-04-21/22~&~12 (6)~& 1600     &1.52-1.00-1.04~&~28-34\\
 
  & 2025-05-04/05~&~18 (9)~& 1200     &1.17-1.00-1.37~&~20-26\\

\hline 
\end{tabular}
\tablefoot{
\tablefoottext{a}{Signal-to-noise ratio (SNR) in the extracted spectrum, per spectral resolution element in the order at 515nm.}}

\end{center}
\end{table*}

\subsection{Rossiter–McLaughlin effect}

The R-M effect causes a spectral line asymmetry, or, equivalently, an asymmetry in the cross-correlation function (CCF) that manifests as anomalous radial velocities during the transit of the exoplanet. To measure the projected spin-orbit angle of the planet ($\lambda$), we followed the methodology of \citet{zak24b, zak25a}: we fitted the RVs with a composite model, which includes a Keplerian orbital component as well as the R-M anomaly. This model is implemented in the \textsc{ARoMEpy}\footnote{\url{https://github.com/esedagha/ARoMEpy}} \citep{seda23} package, which utilizes the \textsc{Radvel} python module \citep{ful18} for the formulation of the Keplerian orbit. \textsc{ARoMEpy} is a Python implementation of the R-M anomaly described in the \textsc{ARoME} code \citep{bou13}. We used the R-M effect function defined for RVs determined through the cross-correlation technique in our code. We set Gaussian priors for the RV semi-amplitude ($K$) and the systemic velocity ($\Gamma$). 

In the R-M effect model, we fixed the following parameters to values reported in the literature: the orbital period ($P$), the planet-to-star radius ratio ($R_{\rm{p}}/R_{\rm{s}}$), and the eccentricity ($e$). The parameter $\sigma$, which is the width of the CCF and represents the effects of the instrumental and turbulent broadening, was measured on the data and fixed. Furthermore, we used the \textit{ExoCTK}\footnote{\url{https://github.com/ExoCTK/exoctk}} tool to compute the quadratic limb-darkening coefficients with ATLAS9 model atmospheres \citep{cas03} in the wavelength range of the PLATOSpec instrument (380-680\,nm). We set Gaussian priors using the literature value and uncertainty derived from transit modeling (Tab.~\ref{tab:targets_compiled}) on the following parameters during the fitting procedure: the central transit time ($T_C$), the orbital inclination ($i$), and the scaled semi-major axis ($a/R_{\rm{s}}$). Gaussian priors were set on the projected stellar rotational velocity ($\nu\,\sin i_*$) and uniform priors on the sky-projected angle between the stellar rotation axis and the normal of the orbital plane ($\lambda$).

To obtain the best fitting values of the parameters, we employed three independent Markov chain Monte Carlo (MCMC) ensemble simulations relying on the \textsc{Infer}\footnote{\url{https://github.com/nealegibson/Infer}} implementation that uses the Affine-Invariant Ensemble Sampler. We initialized the MCMC at parameter values found by the Nelder-Mead method perturbed by a small value. We used 20 walkers each with 12\,500 steps, burning the first 2\,500. As a convergence check, we ensured that the Gelman-Rubin statistics \citep{gel92} is less than 1.001 for each parameter. Using this setup we obtain our results and present them in 
Sect.~\ref{sec:resrm} and in Figs.~\ref{f:35PS} to \ref{f:K2237PS}.

\begin{table*}[htbp]
\caption{Properties of the targets (star and planet).}
\vspace{-.4cm}
\label{tab:targets_compiled}
\begin{center}
\begin{tabular}{@{ }l@{ }l@{ }c@{ }c@{ }c@{ }} 
\hline
& Parameters & WASP-35 & TOI-622 & K2-237 \\ 
\hline
\textbf{Star} & V mag & $10.95 \pm 0.09$& $8.995 \pm 0.002$ & $11.60 \pm 0.05$ \\
 & Sp. Type  & F9V & F6V & F6V \\
 & M$_{\rm{s}}$ (M$_\odot$) & $1.106 \pm 0.015$ & $1.313 \pm 0.079$ & $1.256^{+0.055}_{-0.062}$ \\
 & R$_{\rm{s}}$ (R$_\odot$) & $1.122 \pm 0.016$ & $1.415 \pm 0.047$ & $1.236^{+0.043}_{-0.036}$ \\
 & T$_{\rm{eff}}$ (K) & $6072 \pm 63$ & $6400 \pm 100$ & $6180^{+160}_{-140}$ \\
 & $v\,\rm{sin}i_*$ (km/s) & $2.4 \pm 0.6$ & $19.0 \pm 0.9$ & $11.76 \pm 0.90$ \\
\textbf{Planet} & M$_{\rm{p}}$ (M$_{\rm Jup}$) & $0.765 \pm 0.029$ & $0.303^{+0.069}_{-0.072}$ & $1.366^{+0.11}_{-0.092}$ \\
 & R$_{\rm{p}}$ (R$_{\rm Jup}$) & $1.349 \pm 0.022$ & $0.824^{+0.028}_{-0.029}$ & $1.433^{+0.056}_{-0.049}$ \\
 & Period (d) & $3.1615691 \pm 0.0000003$ & $6.402513^{+0.000031}_{-0.000054}$ & $2.18053332^{+0.00000054}_{-0.00000054}$ \\
 & $\rm{T_0}$ - 2450000 (d) & $5531.47920 \pm 0.00029$ & $8520.69176^{+0.00031}_{-0.00046}$ & $7706.61618^{+0.00003}_{-0.00003}$ \\
 & a (AU) & $0.04360 \pm 0.00020$ & $0.078^{+0.0052}_{-0.0059}$ & $0.03552^{+0.00051}_{-0.00060}$ \\
 & e & $0$ & $0$ & 0 \\
 & i (deg) & $87.95^{+0.33}_{-0.33}$ & $86.62^{+0.77}_{-0.54}$ & $88.37^{+1.0}_{-0.88}$ \\
 & T$_{\rm{eq}}$ (K) & $1484 \pm 18$ & $1388^{+22}_{-22}$ & $1759^{+49}_{-42}$ \\
 & Discovery ref. & \citet{eno11} & \citet{psa23} & \citet{soto18} \\
 & Prior ref. & \citet{bai22} & \citet{psa23} & \citet{thy23} \\
\hline
\end{tabular}
\end{center}
\end{table*}

\section{Projected spin-orbit angles of WASP-35b, TOI-622b and K2-237b}
\label{sec:resrm}

We measured the projected spin-orbit angle ($\lambda$) of three gaseous planets. WASP-35b was previously studied by \citet{zak25b} and we show PLATOSpec data of WASP-35 to display the capabilities of the new instrument. We provide measurements of two targets for the first time: TOI-622b and K2-237b. 

\textbf{WASP-35b} is an inflated hot Jupiter orbiting a F9V host star on a short 3.2-day orbit \citep{eno11}.  In our analysis, we infer an aligned orbit of WASP-35b with $\lambda = 1_{-18}^{+19}$~deg. This is in good agreement with the one measured by \citet{zak25b}, who inferred $\lambda =-5 \pm 11$\,deg using HARPS-N.

Our result comes from two transits observed with the PLATOSpec instrument while the result from \citet{zak25b} uses a single transit with HARPS-N mounted on the 3.58 m Telescope Nazionale Galileo (TNG). The weather was excellent with a sub-arcsec seeing on all three nights. During the first PLATOSpec transit night with 1200s exposures, the RV uncertainties of WASP-35 vary between 11.8 and 20.1 m/s with a median uncertainty of 14.7 m/s. The RV uncertainties from HARPS-N data vary between 3.1 m/s and 9.3 m/s with a median uncertainty of 4.3 m/s and an exposure time of 600s. Factoring in the smaller collecting aperture (factor of 5.6) of the 1.52 m telescope and longer exposure time of the PLATOSpec observations (factor of two), PLATOSpec collected  roughly one-third of the photons of HARPS-N. Since the RV error scales as (S/N)$^{-1}$, the expected RV precision of a HARPS-N-like instrument on the 1.52 m telescope should be 5 - 15 m/s. This speaks well for the performance of PLATOSpec.

We show the comparison between the individual PLATOSpec and HARPS-N data in the Appendix in Fig. \ref{f:35PSH}. Finally, we performed a joint fit using both HARPS-N and PLATOSpec data, yielding $\lambda =-4 \pm 10$\,deg. We show the posterior of the obtained projected spin-orbit angle and projected rotational velocity for each dataset as well as the joint fit in Fig.~\ref{f:35PSHcomp} of the Appendix.

\textbf{TOI-622} is a sub-Jovian planet orbiting a F6V host star on a 6.4-day orbit, as reported by \citet{psa23}, who also noticed that despite the high insolation flux (F$_\oplus$) and adolescent age \citep[$< 1$ Gyr,][]{kou20, psa23}, TOI-622b does not display an inflated radius, contrarily to what is commonly observed in highly irradiated planets. However, the inflated radius' origin and the dependency on the orbital parameters is still not fully understood \citep{tho24}. In our analysis, we infer an aligned orbit for TOI-622b with $\lambda =-4 \pm 12$\,deg.

\textbf{K2-237b} is a hot Jupiter orbiting a F6V host star on a 2.1-day orbit \citep{soto18,smi19}. \citet{shan21} identified significant transit timing variations (TTV) by combining K2 and TESS data. The TTVs were also detected by \citet{yan24} who also derived orbital period decay. We infer a polar orbit with spin-orbit angle $\lambda = 91 \pm 7$~deg.

\begin{figure}[h]
\includegraphics[width=0.45\textwidth]{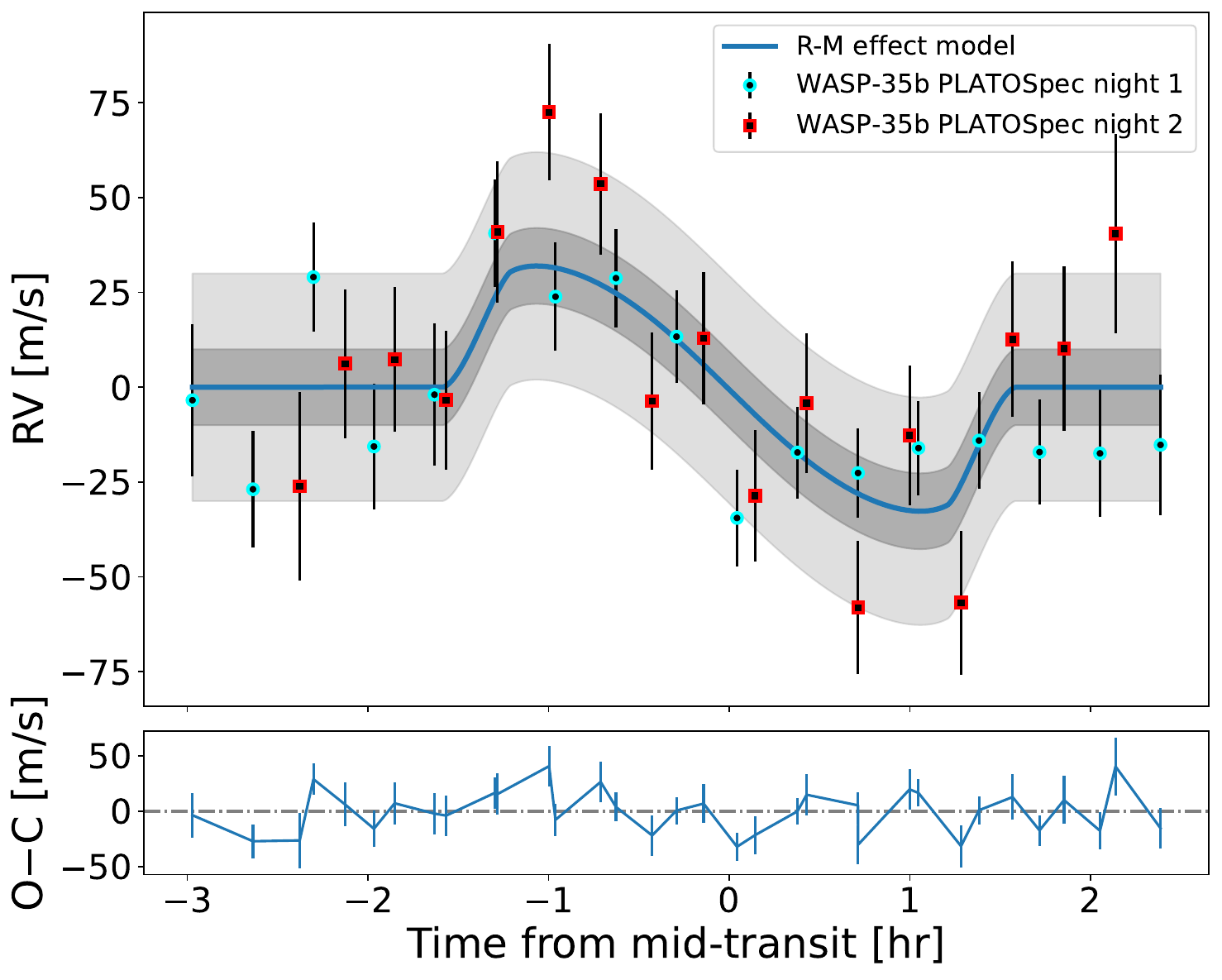}
\caption{R-M effect of WASP-35b observed with PLATOSpec. The observed data points -- colored according to the respective night they were obtained -- are shown with their error bars. The systemic and Keplerian orbit velocities were removed. The blue line shows the best fitting model to the data, together with 1-$\sigma$ (dark grey) and 3-$\sigma$ (light grey) confidence intervals.}
\label{f:35PS}
\end{figure}

\begin{figure}[h]
\includegraphics[width=0.45\textwidth]{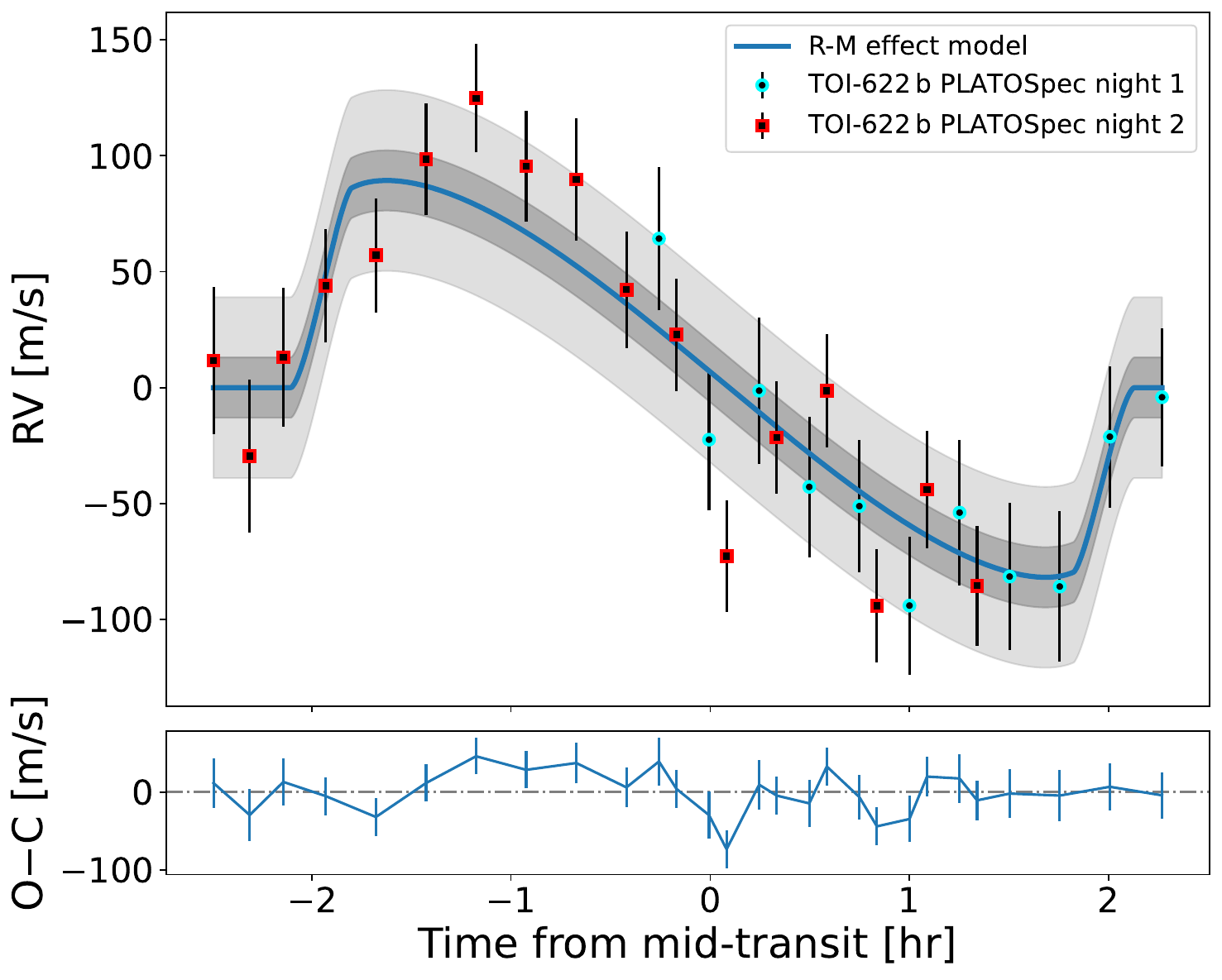}
\caption{Same as Fig. \ref{f:35PS} for TOI-622.}
\label{t622}
\end{figure}

\begin{figure}[h]
\includegraphics[width=0.45\textwidth]{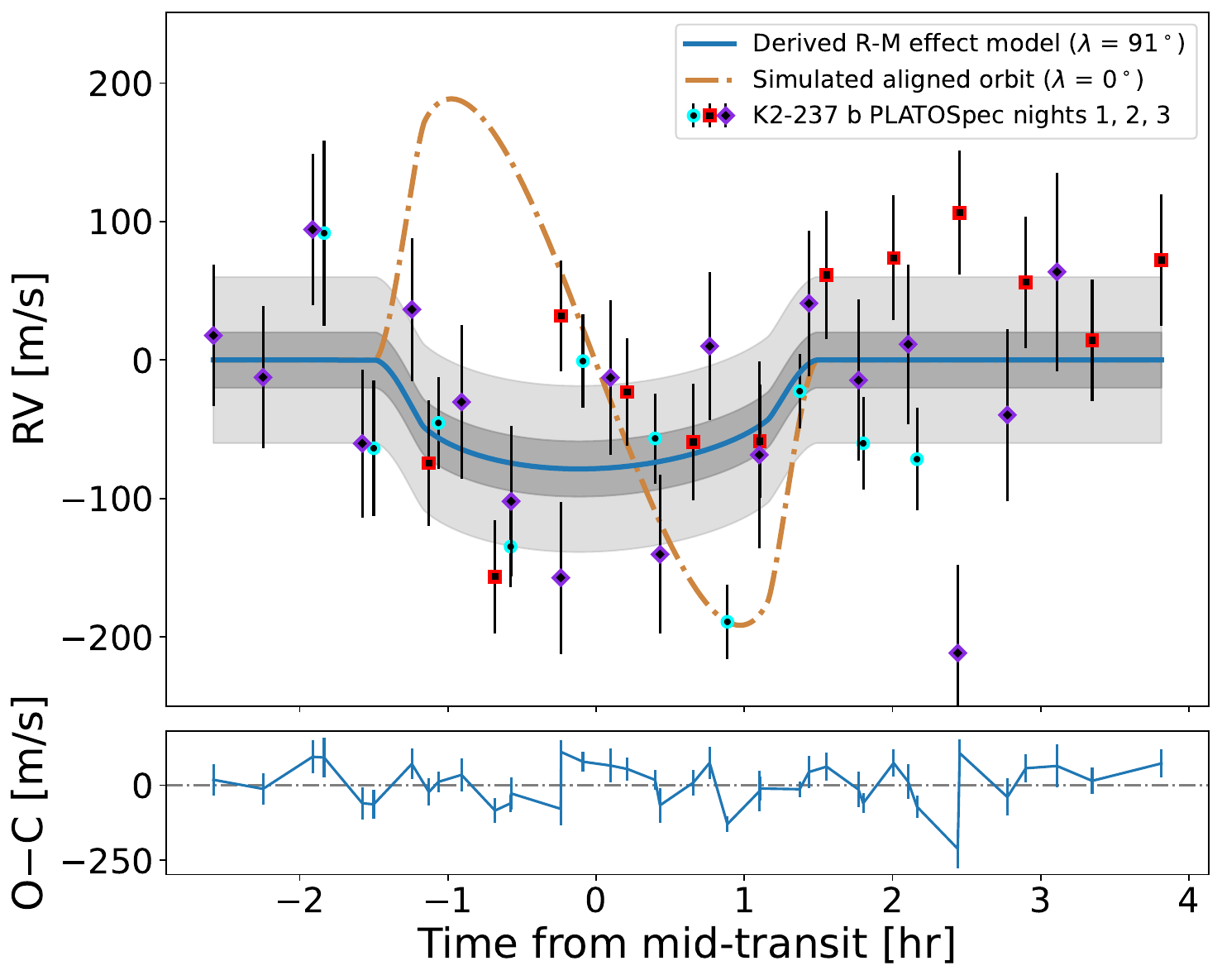}
\caption{Same as Fig. \ref{f:35PS} for K2-237b. Additionally, we show an aligned orbit (brown color) for illustration that is clearly disfavored by the data.}
\label{f:K2237PS}
\end{figure}

The results of our fits are shown in Figs. \ref{f:35PS} to \ref{f:K2237PS}, while the MCMC results are shown in the Appendix in Figs. \ref{f:mcmc35} to \ref{f:mcmck2237}. The derived values from the MCMC analysis are displayed in Table~\ref{tab:resmcmc}.

\begin{table*}[htbp]
\caption{MCMC analysis results for selected exoplanets. $\mathcal{N}$ denotes priors with a normal distribution and $\mathcal{U}$ priors with a uniform distribution.}
\small
\label{tab:resmcmc}
\centering
\begin{tabular}{l | c c | c c | c c}
\hline\hline
 Parameter & Prior & WASP-35b & Prior & TOI-622b & Prior & K2-237b \\
\hline
$T_c$ - 2450000 [d] & $\mathcal{N}(T_0, 0.006)$ & 10681.6770 $\pm$ 0.0005 & $\mathcal{N}(T_0, 0.006)$ & 10729.5639 $\pm$ 0.0038 & $\mathcal{N}(T_0, 0.006)$ & 10752.8178$^{+0.0073}_{-0.0083}$ \\
$\lambda$ [deg] & $\mathcal{U}(-180, 180)$ & 1$^{+18}_{-19}$ & $\mathcal{U}(-180, 180)$ & -4$^{+12}_{-12}$ & $\mathcal{U}(-180, 180)$ & 91 $\pm$ 7 \\
$v\,\sin{i_*}$ [km/s] & $\mathcal{N}(2.4, 0.6)$ & 2.40$^{+0.36}_{-0.35}$ & $\mathcal{N}(19.0, 0.9)$ & 19.4$^{+0.83}_{-0.86}$ & $\mathcal{N}(11.76, 0.90)$ & 11.67$^{+0.90}_{-0.90}$ \\
$a/R_s$ & $\mathcal{N}(8.36, 0.13)$ & 8.32 $\pm$ 0.12 & $\mathcal{N}(10.64, 0.85)$ & 10.97 $^{+0.56}_{-0.58}$ & $\mathcal{N}(6.18, 0.19)$ & 6.16$^{+0.19}_{-0.19}$ \\
$i$ [deg] & $\mathcal{N}(87.95, 0.33)$ & 88.09 $\pm$ 0.32 & $\mathcal{N}(86.62, 0.77)$ & 87.55$^{+0.51}_{-0.52}$ & $\mathcal{N}(88.37, 1.0)$ & 88.11$^{+0.59}_{-0.60}$ \\
$\Gamma$ [km/s] & $\mathcal{N}(17.73, 0.20)$ & $17.754 \pm 0.005$ & $\mathcal{N}(15.80, 0.20)$ & $15.900^{+0.011}_{-0.012}$ & $\mathcal{N}(-22.45, 0.20)$ & $-22.347^{+0.017}_{-0.016}$ \\
$K$ [km/s] & $\mathcal{N}(0.09, 0.05)$ & 0.078 $\pm$ 0.030 & $\mathcal{N}(0.03, 0.05)$ & $0.049^{+0.047}_{-0.048}$& $\mathcal{N}(0.18, 0.10)$ & $0.113^{+0.065}_{-0.062}$ \\
\hline
\end{tabular}
\end{table*}

\section{Discussion}
\subsection{Stellar rotation and true spin-orbit angle $\psi$ }


The projected spin-orbit angle derived through the R-M effect characterizes the architecture with a more accurate description than 
mutual inclinations of planes in multiplanetary systems. However, what is desirable is the true (3D) spin-orbit angle, $\psi$, and not just the projected angle. This requires a knowledge of the stellar inclination angle. 

The stellar inclination can be derived from the stellar rotation period, radius, and projected rotational velocity. One of the best ways to determine the rotation period is by using photometry to study the rotational modulation of the host stars due to, e.g., stellar spots \citep{ska22}. Once the stellar inclination is known, the true spin-orbit angle, $\psi$ (i.e., the angle between the stellar spin-axis and the normal to the orbital plane) can be inferred using the spherical law of cosines, $\cos \psi = \sin i_*\,\sin i\,\cos |\lambda|+\cos i_*\,\cos i$, with the approach suggested by \citet{mas20} to account for the dependency between $v$ and $v\,\sin{i_*}$. A rotational period of $5.07 \pm 0.02$~d was previously determined for K2-237 from K2 mission data \citep{soto18}. We attempted to derive the stellar rotation period and additional variability for WASP-35 and TOI-622 by performing Lomb-Scargle periodograms on the \textit{TESS} \citep{ricker15} data and ASAS-SN \textit{g}-band light curves \citep{ko17} using \textsc{Period04} \citep{Lenz2004}.

\begin{figure}
\centering
\includegraphics[width=0.45\textwidth]{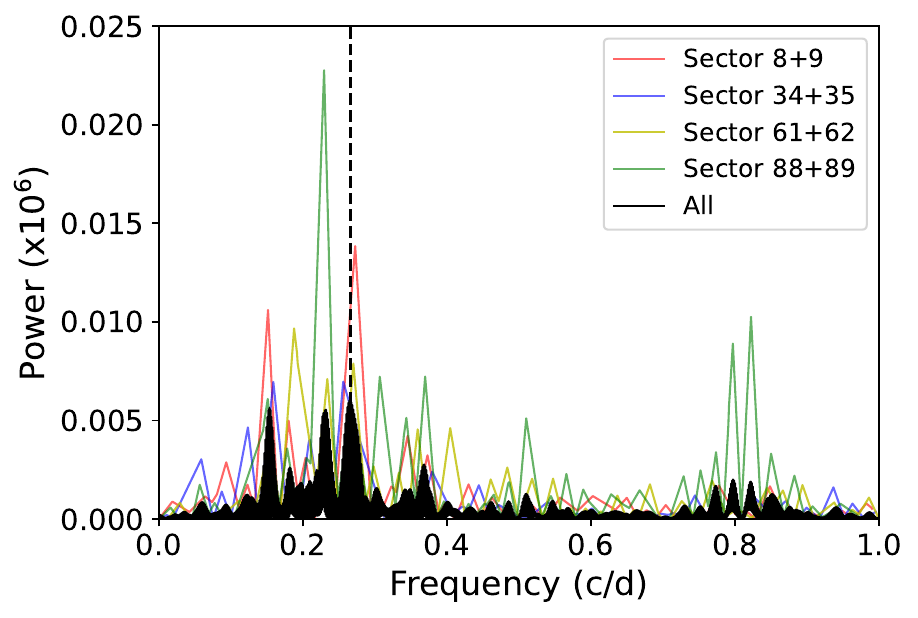}
\caption{TOI-622 frequency spectra of the SPOC \textit{TESS} full data set (black) and from different sectors (colors). The vertical line shows the position of the rotation frequency.}
\label{Fig:FreqSpec}
\end{figure}

We do not find any significant periodic signals for WASP-35. For TOI-622, we investigated data from the \textit{TESS} Science Processing Operations Center (SPOC) and Quick Look pipeline (QLP) pipelines \citep{jen16,hua20a,hua20b}. The data were available in 8 sectors (8+9, 34+35, 61+62 and 88+89) in both long and short cadences (data with a cadence of 1800, 600, 200 and 120 seconds). We investigated all available data sets and excluded the QLP data due to strong artifacts and trends. After removing the transits, we performed a frequency analysis to search for significant periodic variability. According to \citet{psa23}, the rotational period of TOI-622 estimated from $v\,\sin{i_*}$ should be 3.77\,days (0.265\,c/d). This frequency is the strongest one in the \textit{TESS} SPOC 2-minute cadence data and has a \textit{SNR}$=5.1$\footnote{False alarm probability, FAP=$8.2\times10^{-8}$ using equation 7.27 from \citet{hat19}, based on an empirical relation of obtaining FAP from the SNR \citep{kus97}. }, that is above the significance level of \textit{SNR}$=4$ \citep{bre93}. In addition, this peak is apparent in the SPOC long-cadence data and is present in frequency spectra of all sectors when analyzed separately (see Fig.~\ref{Fig:FreqSpec}), although at slightly different positions. This can be a sign of differential rotation. To conclude, we consider the peak at 0.2658\,c/d as the consequence of the rotation of the star. 

Subsequently, using the $v\,\sin{i_*}$ from the R-M effect analysis we calculate $\psi= $16.1$^{+8.0}_{-9.7}$\,deg for TOI-622b that further supports the aligned configuration. Furthermore, we calculate $\psi= $90.5$^{+6.8}_{-6.2}$\,deg for K2-237b. The true spin-orbit measurement confirms the polar orbit for K2-237. In the following sections, we discuss the tidal realignment timescales and possible evolutionary pathways of the studied planets.

\subsection{Dynamical timescale}

To better understand the evolution of these transiting exoplanets we must compare their age to the tidal circularization and tidal realignment timescales. These can be used to study whether the eccentricity and spin-orbit angle have changed since the formation of the system. It provides only a limit as the processes altering these parameters may not happen at the same time as the system's formation.

WASP-35b is most likely the oldest system in our sample as \citet{eno11} derived an age of $6^{+5}_{-4}$ Gyr or it. \citet{psa23} estimated the age of TOI-622 to be $0.9 \pm 0.2$ Gyr, while \citet{thy23} inferred an age of $1.09^{+1.50}_{-0.78}$ Gyr for K2-237.

We have used Equation 3 from \citet{ad06} to calculate the tidal circularization time scale, $ \tau_{\text{cir}}$. Assuming  a tidal quality factor $Q_P=10^6$, we derive $ \tau_{\text{cir}}$ equal to $\approx$ 0.1 Gyr for WASP-35b, $\approx$ 10 Gyr for TOI-622b and $\approx$ 0.02 Gyr for K2-237. We thus conclude that only TOI-622b would likely be able to keep any initial eccentricity, while the other two planets have had their orbits circularized.

Next, we compute the tidal realignment timescales.
For cool stars with convective envelopes below the Kraft break \citep[an abrupt decrease in the stars' average rotation rates at about 6200 K,][]{kraft67}, the tidal alignment timescale can be approximated \citep{alb12} as:

\begin{equation}
\label{ce}
\hspace*{0.25\columnwidth}    \tau_{\rm{CE}} = \frac{10^{10}\,\rm{yr}}{\left( M_{\rm{p}}/M_{\rm{*}} \right)^2}  \left( \frac{a/R_{\rm{*}}}{40} \right)^6.
\end{equation}

\noindent
This allows us to compute the tidal realignment timescale for WASP-35b as $\tau_{\rm{CE}} \approx 10^{12}$\,yr.

Stars above the Kraft break have radiative envelopes with less efficient tidal forces. Hence, a different formula should be used to calculate the tidal realignment timescales. 
TOI-622 has an effective temperature of 6\,400 K and is clearly located above the Kraft break. 
The effective temperature of K2-237 ($6\,257 \pm 100$ K \citep{soto18} and $6\,180^{+160}_{-140}$ \citep{thy23}) is just around the division. However, as presented in \citet{spal22} the division between radiative and convective envelopes is also metallicity dependent (their Fig. 9), and so the derived super-solar metallicity of K2-237 places the K2-237 above the Kraft break.

For these two systems, we used the following formula \citep{alb12} to calculate $\tau_{\rm{RA}}$:

\begin{equation}
\label{ra}
\hspace*{0.08\columnwidth}    \tau_{\rm{RA}} = \frac{5}{4} \cdot 10^9\, \text{yr}  \left( \frac{M_{\rm{p}}}{M_{\rm{*}}} \right)^{-2}  \left(1 + \frac{M_{\rm{p}}}{M_{\rm{*}}}\right)^{-5/6}  \left( \frac{a / R_{\star}}{6} \right)^{17/2},
\end{equation}

\noindent
leading to $\tau_{\rm{RA}} \approx  10^{18}$\,yr for TOI-622b and $\approx  10^{15}$\,yr for K2-237b.

The much longer tidal realignment timescales compared to the age of the studied systems suggest that the spin-orbit angle has not changed since its initial value. This is especially interesting for K2-237b, as it suggests that the planet's polar orbit was established by some initial dynamical events rather than being a later result of tidal realignment.

\subsection{Evolutionary scenarios and spin-orbit distribution}
\textbf{WASP-35b} was studied by \citet{zak25b} with the HARPS-N instrument. They derived an aligned orbit consistent with quiet disc migration. Our result using a smaller, 1.52 m telescope is in good agreement with the previous result (Figs. \ref{f:35PSH} and \ref{f:35PSHcomp}).
\vspace{0.2cm}

\noindent
\textbf{TOI-622b} The low spin-orbit angle of TOI-622b together with zero eccentricity and the long circularization and realignment timescales are indicative of the quiet evolution of the system in the disc, ruling out mechanisms exciting the planetary eccentricity or the spin-orbit angle. 

There is a lack of measurements for Neptunian planets above the Kraft break.
TOI-622 belongs to a rare type of system with an F-type star harboring a sub-Jovian planet. The only other example of a Neptunian planet around an F-type star that confidently lies above the Kraft break with spin-orbit angle constrained is HD\,106315c \citep{zhou18,bou23}. Spin-orbit angle measurement of a sub-Saturn around a more massive star exists, such as TOI-1842b with M$_{\rm{s}}=$ 1.45$^{+0.07}_{-0.14}$ M$_\odot$, for which \citet{hix23} measured $\lambda = -68.1^{+21.2}_{-14.7}$ deg and $\psi = $73.3$^{+16.3}_{-12.9}$ deg. However, the host star is an evolved subgiant with T$_{\rm{eff}}=$6\,033$^{+95}_{-93}$ K, an evolutionary stage for which the Kraft break is no longer applicable. They proposed a hypothesis that hot/massive stars (M$_{\rm{s}}$ $>$ 1.2 M$_{\odot}$) would have a prevalence of misaligned systems. Our result does not fit in this trend.

\begin{figure}
\includegraphics[width=0.49\textwidth]{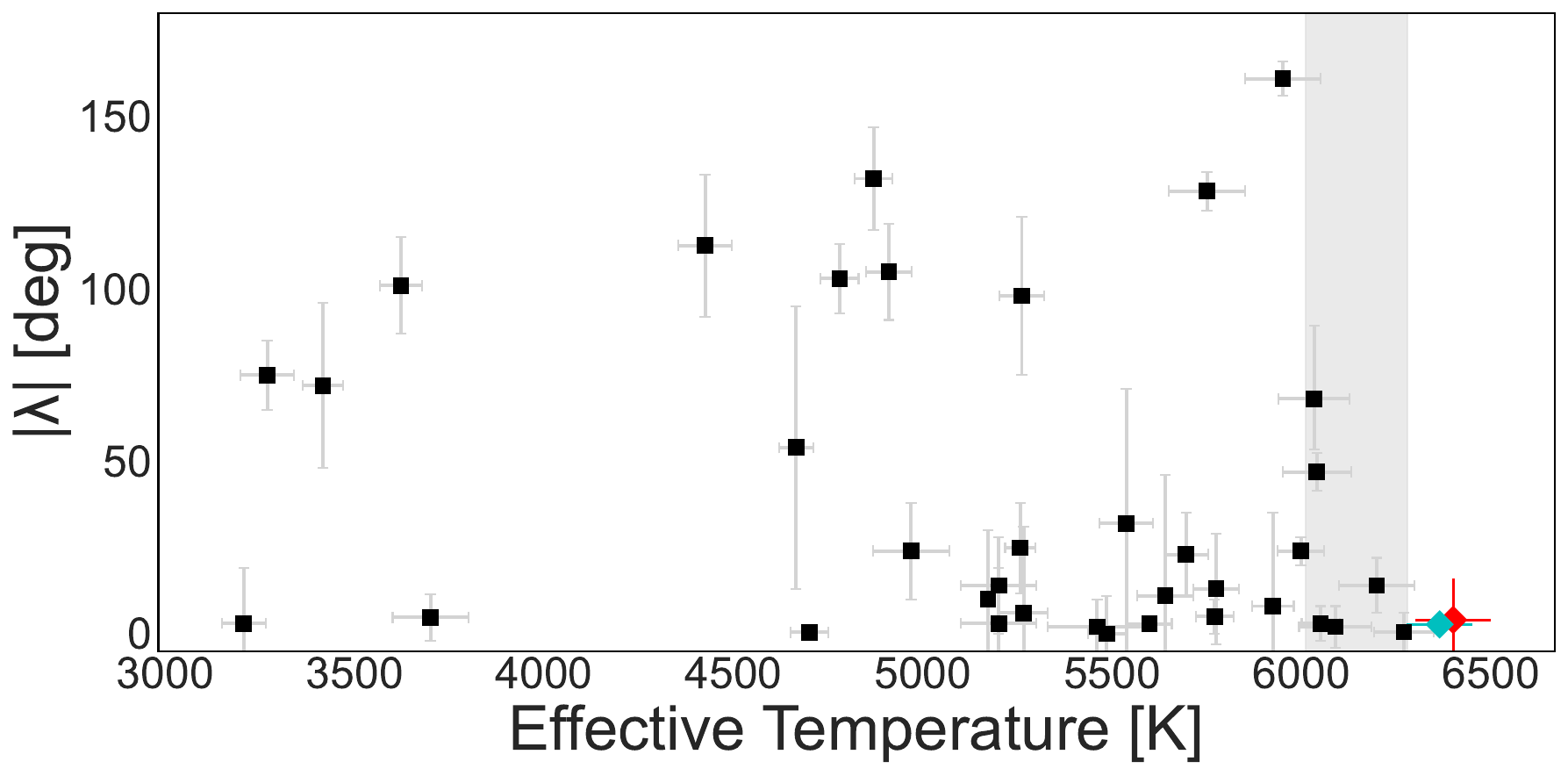}
\caption{Projected spin-orbit angle versus stellar effective temperature for the sub-Jovian population (M$_{\rm{p}}$ $\lesssim$ 0.3 M$_{\rm Jup}$). The gray area shows the position of the Kraft break as derived in \citet{spal22}. The literature values were retrieved from the TEPCat catalogue \citep{south11}. TOI-622b is plotted in red. TOI-622b joins the rare population of sub-Jovian planets around hot stars, and similarly to HD 106315c (cyan), orbits a star located above the Kraft break. }
\label{nep}
\end{figure}

As can be seen from Fig.\,\ref{nep}, more sub-Jovian planets orbiting a star above the Kraft break are needed to explore the shaping mechanisms in this parameter space as it is less affected by tidal realignment. However, finding smaller planets around hotter stars is a challenging prospect. We queried the NASA database to check how many such systems would be available for spin-orbit follow-up.
We imposed conditions of a transiting planet around a star brighter than 12 \textit{V}mag around a star hotter than 6180 K (above the Kraft break) with the following properties: R$_{\rm{p}}$ $\gtrsim$2 R$_\oplus$, M$_{\rm{p}}$ $\lesssim$ 0.3 M$_{\rm Jup}$.
This resulted in 10 systems meeting the conditions that are currently available for a follow-up to determine their spin-orbit angle. A dedicated program could easily achieve measuring the projected spin-orbit angle for most of them with current 4-m and 8-m class telescopes. Deriving their true-spin orbit angles via determination of the stellar inclination might however be challenging as F-type stars display complex behavior where rotation is mixed with various types of pulsations \citep[e.g.,][]{hen23}. Long-term photometric monitoring to fully explain the power spectrum in Fig.~\ref{Fig:FreqSpec} is highly warranted. In this respect, the \textit{PLATO} mission will substantially help, in particular for TOI-622 as it is located in the LOPS2 field \citep{nasc25} that will be observed for at least two continuous years.
\vspace{0.2cm}

\noindent
\textbf{K2-237b} is a typical hot Jupiter. Its host star lies on the boundary between having a convective and a radiative envelope; the former would result in significantly more efficient tidal forces. However, the derived timescales suggest that the spin-orbit angle should not be affected regardless of the inner stellar structure. \citet{yan24} using the detected TTVs and period decay rate suggested that K2-237b has undergone disc migration. We find transit timing consistent with their results, although our uncertainty on the timing does not allow us to further refine the ephemeris. However, in this work we derive a polar orbit that is rather indicative of disc-free migration \citep{fab07} as disc migration is less likely to produce such a highly misaligned orbit. We have checked Gaia DR3 \citep{gaia23} data in search of a companion that could perturb planet's orbit but none of the stars in the proximity are consistent with a physically bound companion. Furthermore, the Renormalised Unit Weight Error (RUWE)\footnote{This parameter captures the astrometric deviations from single-source solutions. See \url{https://gea.esac.esa.int/archive/documentation/GEDR3/Gaia_archive/chap_datamodel/sec_dm_main_tables/ssec_dm_gaia_source.html\#gaia_source-ruwe} for a detailed definition. A value close to 1 indicates that the astrometric solution is valid.} value $\approx$ 0.98 does not suggest binarity. Without the presence of a bound companion, we cannot unequivocally rule out disc migration.

The K2-237 system adds another piece of evidence for the ``Preponderance of Perpendicular Planets" \citep{alb21} suggesting some orbital configurations are preferred. This trend -- if indeed real \citep{dong23, sie23} -- is still not understood and neither are the mechanisms shaping it \citep{att23,knud24}. Such a prevalence has been suggested both for Jupiters and Neptunians, yet the mechanisms responsible for their origin seem to be different. As detailed in \citet{pet20}, the disc-driven resonances proposed to alter the spin-orbit angle of Neptunian planets would be inefficient for more massive planets.

Despite its polar orbit, disc-free migration for K2-237b is not unambiguously confirmed. An alternative explanation may lie in the system's early evolution. The traditional picture assumes a single star and its disc inherit their spin orientation from the initial collapse of a prestellar core and that this alignment is preserved. However, this idealized view often overlooks that the system's initial conditions can be altered by subsequent evolution and interactions with the surrounding environment. Early planet formation is not fully understood, and processes that warp or tilt the disc---potentially caused by stellar magnetic fields, fly-by encounters, or other phenomena---might shape exoplanet populations in ways not yet fully explored \citep{fou11,ro13,bat12,mat17}. One such environmental mechanism capable of breaking the initial stellar-disc alignment is the late infall of material onto the system \citep{bate10,th11,fielding15,kuff21}.


\citet{kuff24} have shown that the infall might be capable of creating polar orbits such as the one of K2-237b.
Additionally, they have found that the misalignment due to infall correlates with stellar mass. Higher mass stars take longer to settle into their final spin orientation compared to lower mass stars. During this settling, the late infall can occur, more easily producing misaligned planets. The combined effect of tidal realignment, which is more efficient for stars below the Kraft break (with convective envelopes), and late infall will produce a distribution of exoplanets with misaligned planets preferentially around stars above the Kraft break ($\gtrsim 1.2$ M$_\odot$), although on vastly different timescales. A promising way to distinguish the dominant mechanism would be to explore the spin-orbit of young planets where the tidal realignment did not have sufficient time to act upon, while the late-infall would have already concluded, alongside the disappearance of the protoplanetary disc. The recent discovery of a 3 Myr old transiting exoplanet with a misaligned disc and the presence of a stellar companion \citep{bar24} support the complex interplay of several mechanisms forming the spin-orbit distribution \citep{nea25}. 


The true origin of the evolution of K2-237b will be unlocked by obtaining additional data: i) searching for the additional companion responsible for the TTVs or one that would be able to produce misaligned orbit by gravitational interaction (Kozai-Lidov mechanism) ii) allowing for detailed atmospheric composition constraining the accretion mechanisms onto the planet during the disc stage. The elemental abundance ratios such as C/O, N/O, S/N \citep{ob11,tur21,pace22} are indicative of previous migration mechanisms. For example, the detection of a high C/O ratio with the JWST or the upcoming \textit{Ariel} mission would serve as additional evidence of the disc-free migration scenario. While a low C/O ratio would rather be indicative of disc migration where the planet accretes oxygen-rich solids that alter the initial composition.

Finally, obtaining the spin-orbit angle measurements is highly complementary to atmospheric studies using space missions such as \textit{JWST} and \textit{Ariel} as planetary migration is expected to alter both the spin-orbit angle and chemical composition \citep{mad14}. However, their exact interplay is not fully understood and should be performed at the population level \citep{kirk24}.
\citet{zak25b} identified a lack of spin-orbit angle measurements within the \textit{Ariel} Mission candidate sample (MCS) list, as over 70\% of candidates do not have their spin-orbit measured. Within Tier 3 candidates (those that will be selected for detailed atmospheric studies over multiple visits) this was only 30\%. 

All three targets in this work are listed on the \textit{Ariel} MCS list among Tier 3 candidates \citep{ed22} and our results can serve as an additional parameter for the final target selection as well as for the subsequent interpretation of the measured atmospheric composition showcasing the synergy between ground- and space-based observations.  

\section{Summary}
The migration pathways of gaseous planets, together with the mechanisms shaping the spin-orbit angle distribution, are still unconstrained. We present measurements of the projected spin-orbit angle of three gaseous exoplanets. We have used the newly commissioned PLATOSpec instrument mounted at the E152 Telescope at La Silla Observatory. We measure the projected spin-orbit angle for WASP-35b to demonstrate the capability of PLATOSpec and find excellent agreement with HARPS-N literature data. Furthermore, we measure the projected spin-orbit angle of TOI-622b and K2-237b for the first time. We derive an aligned orbit for TOI-622b and a nearly perfect polar orbit for K2-237b. We also derive the true spin-orbit angle $\psi$ for TOI-622 and K2-237b. We infer that the previous evolution of TOI-622b is consistent with quiet disc migration. In contrast to previous work, we suggest that K2-237b has likely undergone disc-free migration. Both systems are valuable pieces in the spin-orbit distribution and the quest for its understanding. TOI-622b is located in the sparsely populated region of sub-Jovian planets with spin-orbit angle measurement orbiting stars above the Kraft break. More data are needed to characterize the spin-orbit distribution of sub-Jovian planets around hot stars and how it differs from the one of Jovian planets. K2-237b's polar orbit strengthens the tentative existence of preferential orientations of planetary orbits. Furthermore, all three targets are prime targets for future atmospheric characterization. Finally, our results highlight the capability of the PLATOSpec as a successful instrument that is able to provide ground-based support to \textit{PLATO}, \textit{Ariel}, and other space missions in years to come.

\begin{acknowledgements} The authors thank the anonymous referee for helpful comments that improved the quality of the manuscript. JZ and PK acknowledge support from GACR:22-30516K. The research of PG was supported by the Slovak Research and Development Agency under contract No. APVV-20-0148 and the internal grant No. VVGS-2023-2784 of the P. J. {\v S}af{\'a}rik University in Ko{\v s}ice funded by the EU NextGenerationEU through the Recovery and Resilience Plan for Slovakia under the project No. 09I03-03-V05-00008. LV acknowledges the support from ANID Fondecyt n. 1211162, Fondecyt n.
1251299, and BASAL FB210003. 
AB is supported by the Italian Space Agency (ASI) with \textit{Ariel} grant n. 2021.5.HH.0. 
RB acknowledges support from \textsc{fondecyt} Project 1241963 and from \textsc{anid} -- Millennium  Science  Initiative -- ICN12\_009.
DI acknowledges support from collaborations and/or information exchange within NASA’s Nexus for Exoplanet System Science (NExSS) research coordination network sponsored by NASA’s Science Mission Directorate under
agreement no. 80NSSC21K0593 for the programme ‘Alien Earths’. PLATOSpec was built and is operated by a consortium consisting of the
Astronomical Institute ASCR in Ondrejov, Czech Republic (ASU), the
Thüringer Landessternwarte (Thuringian State Observatory - Germany), the
Universidad Catholica in Chile (PUC - Chile), and minor partners include
Masaryk University (Czechia), Universidad Adolfo Ibanez (Chile) and
Institute for PLasma Physics of the Czech Academy of Sciences (Czechia).
Financing for the modernisation and front end of the 1.52-m telescope
was provided by AsU and personal costs were partly financed from grant
LTT-20015. Financing for the construction of PLATOSpec was provided by
the Free State of Thuringia, under the "Directive for the Promotion of
Research PUC is acknowledging the support from ANID Fondecyt n. 1211162
and n. 1251299, and ANID QUIMAL ASTRO20-0025. Use of the 1.52-m
telescope was made possible through an agreement between ESO and the
PLATOSpec consortium. 
\end{acknowledgements}

\bibliographystyle{aa}
\bibliography{aa}

\begin{thebibliography}{88}
\expandafter\ifx\csname natexlab\endcsname\relax\def\natexlab#1{#1}\fi

\bibitem[{{Adams} \& {Laughlin}(2006)}]{ad06}
{Adams}, F.~C. \& {Laughlin}, G. 2006, \apj, 649, 1004

\bibitem[{{Albrecht} {et~al.}(2012){Albrecht}, {Winn}, {Johnson}, \& et~al.}]{alb12}
{Albrecht}, S., {Winn}, J.~N., {Johnson}, J.~A., \& et~al. 2012, \apj, 757, 18

\bibitem[{{Albrecht} {et~al.}(2022){Albrecht}, {Dawson}, \& {Winn}}]{alb22}
{Albrecht}, S.~H., {Dawson}, R.~I., \& {Winn}, J.~N. 2022, \pasp, 134, 082001

\bibitem[{{Albrecht} {et~al.}(2021){Albrecht}, {Marcussen}, {Winn}, \& et~al.}]{alb21}
{Albrecht}, S.~H., {Marcussen}, M.~L., {Winn}, J.~N., \& et~al. 2021, \apjl, 916, L1

\bibitem[{{Attia} {et~al.}(2023){Attia}, {Bourrier}, {Delisle}, \& et~al.}]{att23}
{Attia}, O., {Bourrier}, V., {Delisle}, J.~B., \& et~al. 2023, \aap, 674, A120

\bibitem[{{Bai} {et~al.}(2022){Bai}, {Gu}, {Wang}, {Sun}, {Kwok}, \& {Hui}}]{bai22}
{Bai}, L., {Gu}, S., {Wang}, X., {et~al.} 2022, \aj, 163, 208

\bibitem[{{Bailey} {et~al.}(2016){Bailey}, {Batygin}, \& {Brown}}]{bai16}
{Bailey}, E., {Batygin}, K., \& {Brown}, M.~E. 2016, \aj, 152, 126

\bibitem[{{Barber} {et~al.}(2024){Barber}, {Mann}, {Vanderburg}, \& et~al.}]{bar24}
{Barber}, M.~G., {Mann}, A.~W., {Vanderburg}, A., \& et~al. 2024, \nat, 635, 574

\bibitem[{{Baruteau} {et~al.}(2014){Baruteau}, {Crida}, {Paardekooper}, {Masset}, {Guilet}, {Bitsch}, {Nelson}, {Kley}, \& {Papaloizou}}]{baru14}
{Baruteau}, C., {Crida}, A., {Paardekooper}, S.~J., {et~al.} 2014, in Protostars and Planets VI, ed. H.~{Beuther}, R.~S. {Klessen}, C.~P. {Dullemond}, \& T.~{Henning}, 667--689

\bibitem[{{Bate} {et~al.}(2010){Bate}, {Lodato}, \& {Pringle}}]{bate10}
{Bate}, M.~R., {Lodato}, G., \& {Pringle}, J.~E. 2010, \mnras, 401, 1505

\bibitem[{{Batygin}(2012)}]{bat12}
{Batygin}, K. 2012, \nat, 491, 418

\bibitem[{{Beck} \& {Giles}(2005)}]{bec05}
{Beck}, J.~G. \& {Giles}, P. 2005, \apjl, 621, L153

\bibitem[{{Bieryla} {et~al.}(2025){Bieryla}, {Dong}, {Zhou}, {Eastman}, {Mayorga}, {Latham}, {Carter}, {Huang}, {Quinn}, {Collins}, {Abe}, {Beletsky}, {Brahm}, {Col{\'o}n}, {Essack}, {Guillot}, {Henning}, {Hobson}, {Horne}, {Jenkins}, {Jones}, {Jord{\'a}n}, {Osip}, {Ricker}, {Rodriguez}, {Schulte}, {Schwarz}, {Seager}, {Shporer}, {Suarez}, {Tan}, {Ting}, {Triaud}, {Vanderburg}, {Villase{\~n}or}, {Vowell}, {Watkins}, {Winn}, \& {Ziegler}}]{bie25}
{Bieryla}, A., {Dong}, J., {Zhou}, G., {et~al.} 2025, \aj, 169, 273

\bibitem[{{Bou{\'e}} {et~al.}(2013){Bou{\'e}}, {Montalto}, {Boisse}, \& et~al.}]{bou13}
{Bou{\'e}}, G., {Montalto}, M., {Boisse}, I., \& et~al. 2013, \aap, 550, A53

\bibitem[{{Bourrier} {et~al.}(2023){Bourrier}, {Attia}, {Mallonn}, \& et~al.}]{bou23}
{Bourrier}, V., {Attia}, O., {Mallonn}, M., \& et~al. 2023, \aap, 669, A63

\bibitem[{{Brahm} {et~al.}(2017){Brahm}, {Jord{\'a}n}, \& {Espinoza}}]{brahm17}
{Brahm}, R., {Jord{\'a}n}, A., \& {Espinoza}, N. 2017, \pasp, 129, 034002

\bibitem[{{Breger} {et~al.}(1993){Breger}, {Stich}, {Garrido}, {Martin}, {Jiang}, {Li}, {Hube}, {Ostermann}, {Paparo}, \& {Scheck}}]{bre93}
{Breger}, M., {Stich}, J., {Garrido}, R., {et~al.} 1993, \aap, 271, 482

\bibitem[{{Castelli} \& {Kurucz}(2003)}]{cas03}
{Castelli}, F. \& {Kurucz}, R.~L. 2003, in Modelling of Stellar Atmospheres, ed. N.~{Piskunov}, W.~W. {Weiss}, \& D.~F. {Gray}, Vol. 210, A20

\bibitem[{{Chen} {et~al.}(2024){Chen}, {Baronett}, {Nixon}, \& {Martin}}]{chen24}
{Chen}, C., {Baronett}, S.~A., {Nixon}, C.~J., \& {Martin}, R.~G. 2024, \mnras, 533, L37

\bibitem[{{Crida} \& {Batygin}(2014)}]{cri14}
{Crida}, A. \& {Batygin}, K. 2014, \aap, 567, A42

\bibitem[{{Dawson} \& {Johnson}(2018)}]{daw18}
{Dawson}, R.~I. \& {Johnson}, J.~A. 2018, \araa, 56, 175

\bibitem[{{Dong} \& {Foreman-Mackey}(2023)}]{dong23}
{Dong}, J. \& {Foreman-Mackey}, D. 2023, \aj, 166, 112

\bibitem[{{Edwards} \& {Tinetti}(2022)}]{ed22}
{Edwards}, B. \& {Tinetti}, G. 2022, \aj, 164, 15

\bibitem[{{Enoch} {et~al.}(2011){Enoch}, {Anderson}, {Barros}, \& et~al.}]{eno11}
{Enoch}, B., {Anderson}, D.~R., {Barros}, S.~C.~C., \& et~al. 2011, \aj, 142, 86

\bibitem[{{Espinoza-Retamal} {et~al.}(2023){Espinoza-Retamal}, {Brahm}, {Petrovich}, {Jord{\'a}n}, {Stef{\'a}nsson}, {Sedaghati}, {Hobson}, {Mu{\~n}oz}, {Boyle}, {Leiva}, \& {Suc}}]{esp23}
{Espinoza-Retamal}, J.~I., {Brahm}, R., {Petrovich}, C., {et~al.} 2023, \apjl, 958, L20

\bibitem[{{Espinoza-Retamal} {et~al.}(2024){Espinoza-Retamal}, {Stef{\'a}nsson}, {Petrovich}, {Brahm}, {Jord{\'a}n}, {Sedaghati}, {Lucero}, {Tala Pinto}, {Mu{\~n}oz}, {Boyle}, {Leiva}, \& {Suc}}]{esp24nep}
{Espinoza-Retamal}, J.~I., {Stef{\'a}nsson}, G., {Petrovich}, C., {et~al.} 2024, \aj, 168, 185

\bibitem[{{Fabrycky} \& {Tremaine}(2007)}]{fab07}
{Fabrycky}, D. \& {Tremaine}, S. 2007, \apj, 669, 1298

\bibitem[{{Fielding} {et~al.}(2015){Fielding}, {McKee}, {Socrates}, {Cunningham}, \& {Klein}}]{fielding15}
{Fielding}, D.~B., {McKee}, C.~F., {Socrates}, A., {Cunningham}, A.~J., \& {Klein}, R.~I. 2015, \mnras, 450, 3306

\bibitem[{{Foucart} \& {Lai}(2011)}]{fou11}
{Foucart}, F. \& {Lai}, D. 2011, \mnras, 412, 2799

\bibitem[{{Fulton} {et~al.}(2018){Fulton}, {Petigura}, {Blunt}, \& et~al.}]{ful18}
{Fulton}, B.~J., {Petigura}, E.~A., {Blunt}, S., \& et~al. 2018, \pasp, 130, 044504

\bibitem[{{Gaia Collaboration} {et~al.}(2023){Gaia Collaboration}, {Vallenari}, {Brown}, \& et~al.}]{gaia23}
{Gaia Collaboration}, {Vallenari}, A., {Brown}, A.~G.~A., \& et~al. 2023, \aap, 674, A1

\bibitem[{{Gelman} \& {Rubin}(1992)}]{gel92}
{Gelman}, A. \& {Rubin}, D.~B. 1992, Statistical Science, 7, 457

\bibitem[{{Hatzes}(2019)}]{hat19}
{Hatzes}, A.~P. 2019, {The Doppler Method for the Detection of Exoplanets}

\bibitem[{{Henriksen} {et~al.}(2023){Henriksen}, {Antoci}, {Saio}, \& et~al.}]{hen23}
{Henriksen}, A.~I., {Antoci}, V., {Saio}, H., \& et~al. 2023, \mnras, 524, 4196

\bibitem[{{Hixenbaugh} {et~al.}(2023){Hixenbaugh}, {Wang}, {Rice}, \& {Wang}}]{hix23}
{Hixenbaugh}, K., {Wang}, X.-Y., {Rice}, M., \& {Wang}, S. 2023, \apjl, 949, L35

\bibitem[{{Huang} {et~al.}(2020{\natexlab{a}}){Huang}, {Vanderburg}, {P{\'a}l}, {Sha}, {Yu}, {Fong}, {Fausnaugh}, {Shporer}, {Guerrero}, {Vanderspek}, \& {Ricker}}]{hua20a}
{Huang}, C.~X., {Vanderburg}, A., {P{\'a}l}, A., {et~al.} 2020{\natexlab{a}}, Research Notes of the American Astronomical Society, 4, 204

\bibitem[{{Huang} {et~al.}(2020{\natexlab{b}}){Huang}, {Vanderburg}, {P{\'a}l}, {Sha}, {Yu}, {Fong}, {Fausnaugh}, {Shporer}, {Guerrero}, {Vanderspek}, \& {Ricker}}]{hua20b}
{Huang}, C.~X., {Vanderburg}, A., {P{\'a}l}, A., {et~al.} 2020{\natexlab{b}}, Research Notes of the American Astronomical Society, 4, 206

\bibitem[{{Ida} \& {Lin}(2004)}]{ida04}
{Ida}, S. \& {Lin}, D.~N.~C. 2004, \apj, 604, 388

\bibitem[{{Jenkins} {et~al.}(2016){Jenkins}, {Twicken}, {McCauliff}, {Campbell}, {Sanderfer}, {Lung}, {Mansouri-Samani}, {Girouard}, {Tenenbaum}, {Klaus}, {Smith}, {Caldwell}, {Chacon}, {Henze}, {Heiges}, {Latham}, {Morgan}, {Swade}, {Rinehart}, \& {Vanderspek}}]{jen16}
{Jenkins}, J.~M., {Twicken}, J.~D., {McCauliff}, S., {et~al.} 2016, in Society of Photo-Optical Instrumentation Engineers (SPIE) Conference Series, Vol. 9913, Software and Cyberinfrastructure for Astronomy IV, ed. G.~{Chiozzi} \& J.~C. {Guzman}, 99133E

\bibitem[{{Kabath} {et~al.}(2022){Kabath}, {Vanzi}, {Hatzes}, {Guenther}, {Brahm}, {Janik}, {Minezaki}, {Skarka}, \& {Karjalainen}}]{kab22ps}
{Kabath}, P., {Vanzi}, L., {Hatzes}, A., {et~al.} 2022, in Bulletin of the American Astronomical Society, Vol.~54, 102.118

\bibitem[{{Kirk} {et~al.}(2024){Kirk}, {Ahrer}, {Penzlin}, \& et~al.}]{kirk24}
{Kirk}, J., {Ahrer}, E.-M., {Penzlin}, A. B.~T., \& et~al. 2024, RAS Techniques and Instruments, 3, 691

\bibitem[{{Knudstrup} {et~al.}(2024){Knudstrup}, {Albrecht}, {Winn}, {Gandolfi}, {Zanazzi}, {Persson}, {Fridlund}, {Marcussen}, {Chontos}, {Keniger}, {Eisner}, {Bieryla}, {Isaacson}, {Howard}, {Hirsch}, {Murgas}, {Narita}, {Palle}, {Kawai}, \& {Baker}}]{knud24}
{Knudstrup}, E., {Albrecht}, S.~H., {Winn}, J.~N., {et~al.} 2024, \aap, 690, A379

\bibitem[{{Kochanek} {et~al.}(2017){Kochanek}, {Shappee}, {Stanek}, {Holoien}, {Thompson}, {Prieto}, {Dong}, {Shields}, {Will}, {Britt}, {Perzanowski}, \& {Pojma{\'n}ski}}]{ko17}
{Kochanek}, C.~S., {Shappee}, B.~J., {Stanek}, K.~Z., {et~al.} 2017, \pasp, 129, 104502

\bibitem[{{Kounkel} {et~al.}(2020){Kounkel}, {Covey}, \& {Stassun}}]{kou20}
{Kounkel}, M., {Covey}, K., \& {Stassun}, K.~G. 2020, \aj, 160, 279

\bibitem[{{Kraft}(1967)}]{kraft67}
{Kraft}, R.~P. 1967, \apj, 150, 551

\bibitem[{{Kuffmeier} {et~al.}(2021){Kuffmeier}, {Dullemond}, {Reissl}, \& {Goicovic}}]{kuff21}
{Kuffmeier}, M., {Dullemond}, C.~P., {Reissl}, S., \& {Goicovic}, F.~G. 2021, \aap, 656, A161

\bibitem[{{Kuffmeier} {et~al.}(2024){Kuffmeier}, {Pineda}, {Segura-Cox}, \& {Haugb{\o}lle}}]{kuff24}
{Kuffmeier}, M., {Pineda}, J.~E., {Segura-Cox}, D., \& {Haugb{\o}lle}, T. 2024, \aap, 690, A297

\bibitem[{{Kuschnig} {et~al.}(1997){Kuschnig}, {Weiss}, {Gruber}, {Bely}, \& {Jenkner}}]{kus97}
{Kuschnig}, R., {Weiss}, W.~W., {Gruber}, R., {Bely}, P.~Y., \& {Jenkner}, H. 1997, \aap, 328, 544

\bibitem[{{Lai}(2012)}]{lai12}
{Lai}, D. 2012, \mnras, 423, 486

\bibitem[{{Lenz} \& {Breger}(2004)}]{Lenz2004}
{Lenz}, P. \& {Breger}, M. 2004, in IAU Symposium, Vol. 224, The A-Star Puzzle, ed. J.~{Zverko}, J.~{Ziznovsky}, S.~J. {Adelman}, \& W.~W. {Weiss}, 786--790

\bibitem[{{Louden} \& {Millholland}(2024)}]{lo24}
{Louden}, E.~M. \& {Millholland}, S.~C. 2024, \apj, 974, 304

\bibitem[{{Madhusudhan} {et~al.}(2016){Madhusudhan}, {Ag{\'u}ndez}, {Moses}, \& {Hu}}]{madhu16}
{Madhusudhan}, N., {Ag{\'u}ndez}, M., {Moses}, J.~I., \& {Hu}, Y. 2016, \ssr, 205, 285

\bibitem[{{Madhusudhan} {et~al.}(2014){Madhusudhan}, {Amin}, \& {Kennedy}}]{mad14}
{Madhusudhan}, N., {Amin}, M.~A., \& {Kennedy}, G.~M. 2014, \apjl, 794, L12

\bibitem[{{Masuda} \& {Winn}(2020)}]{mas20}
{Masuda}, K. \& {Winn}, J.~N. 2020, \aj, 159, 81

\bibitem[{{Matsakos} \& {K{\"o}nigl}(2017)}]{mat17}
{Matsakos}, T. \& {K{\"o}nigl}, A. 2017, \aj, 153, 60

\bibitem[{{McLaughlin}(1924)}]{mcl}
{McLaughlin}, D.~B. 1924, \apj, 60, 22

\bibitem[{{Nascimbeni} {et~al.}(2025){Nascimbeni}, {Piotto}, {Cabrera}, {Montalto}, {Marinoni}, {Marrese}, {Aerts}, {Altavilla}, {Benatti}, {B{\"o}rner}, {Deleuil}, {Desidera}, {Gizon}, {Goupil}, {Granata}, {Heras}, {Magrin}, {Malavolta}, {Mas-Hesse}, {Osborn}, {Pagano}, {Paproth}, {Pollacco}, {Prisinzano}, {Ragazzoni}, {Ramsay}, {Rauer}, {Tkachenko}, \& {Udry}}]{nasc25}
{Nascimbeni}, V., {Piotto}, G., {Cabrera}, J., {et~al.} 2025, \aap, 694, A313

\bibitem[{{Nealon} {et~al.}(2025){Nealon}, {Smallwood}, {Aly}, {Winter}, {Longarini}, {Cuello}, {Veras}, \& {Alexander}}]{nea25}
{Nealon}, R., {Smallwood}, J.~L., {Aly}, H., {et~al.} 2025, \mnras, 540, L84

\bibitem[{{{\"O}berg} {et~al.}(2011){{\"O}berg}, {Murray-Clay}, \& {Bergin}}]{ob11}
{{\"O}berg}, K.~I., {Murray-Clay}, R., \& {Bergin}, E.~A. 2011, \apjl, 743, L16

\bibitem[{{Pacetti} {et~al.}(2022){Pacetti}, {Turrini}, {Schisano}, \& et~al.}]{pace22}
{Pacetti}, E., {Turrini}, D., {Schisano}, E., \& et~al. 2022, \apj, 937, 36

\bibitem[{{Petrovich} {et~al.}(2020){Petrovich}, {Mu{\~n}oz}, {Kratter}, \& et~al.}]{pet20}
{Petrovich}, C., {Mu{\~n}oz}, D.~J., {Kratter}, K.~M., \& et~al. 2020, \apjl, 902, L5

\bibitem[{{Psaridi} {et~al.}(2023){Psaridi}, {Bouchy}, {Lendl}, {Akinsanmi}, {Stassun}, {Smalley}, {Armstrong}, {Howard}, {Ulmer-Moll}, {Grieves}, {Barkaoui}, {Rodriguez}, {Bryant}, {Su{\'a}rez}, {Guillot}, {Evans}, {Attia}, {Wittenmyer}, {Yee}, {Collins}, {Zhou}, {Galland}, {Parc}, {Udry}, {Figueira}, {Ziegler}, {Mordasini}, {Winn}, {Seager}, {Jenkins}, {Twicken}, {Brahm}, {Jones}, {Abe}, {Addison}, {Brice{\~n}o}, {Briegal}, {Collins}, {Daylan}, {Eigm{\"u}ller}, {Furesz}, {Guerrero}, {Hagelberg}, {Heitzmann}, {Hounsell}, {Huang}, {Krenn}, {Law}, {Mann}, {McCormac}, {M{\'e}karnia}, {Mounzer}, {Nielsen}, {Osborn}, {Reinarz}, {Sefako}, {Steiner}, {Str{\o}m}, {Triaud}, {Vanderspek}, {Vanzi}, {Vines}, {Watson}, {Wright}, \& {Zapata}}]{psa23}
{Psaridi}, A., {Bouchy}, F., {Lendl}, M., {et~al.} 2023, \aap, 675, A39

\bibitem[{{Rauer} {et~al.}(2025){Rauer}, {Aerts}, {Cabrera}, {Deleuil}, {Erikson}, {Gizon}, {Goupil}, {Heras}, {Walloschek}, {Lorenzo-Alvarez}, {Marliani}, {Martin-Garcia}, {Mas-Hesse}, {O'Rourke}, {Osborn}, {Pagano}, {Piotto}, {Pollacco}, {Ragazzoni}, {Ramsay}, {Udry}, {Appourchaux}, {Benz}, {Brandeker}, {G{\"u}del}, {Janot-Pacheco}, {Kabath}, {Kjeldsen}, {Min}, {Santos}, {Smith}, {Suarez}, {Werner}, {Aboudan}, {Abreu}, {Acu{\~n}a}, {Adams}, {Adibekyan}, {Affer}, {Agneray}, {Agnor}, {Aguirre B{\o}rsen-Koch}, {Ahmed}, {Aigrain}, {Al-Bahlawan}, {Alcacera Gil}, {Alei}, {Alencar}, {Alexander}, {Alfonso-Garz{\'o}n}, {Alibert}, {Allende Prieto}, {Almeida}, {Alonso Sobrino}, {Altavilla}, {Althaus}, {Alvarez Trujillo}, {Amarsi}, {Ammler-von Eiff}, {Am{\^o}res}, {Andrade}, {Antoniadis-Karnavas}, {Ant{\'o}nio}, {Aparicio del Moral}, {Appolloni}, {Arena}, {Armstrong}, {Aroca Aliaga}, {Asplund}, {Audenaert}, {Auricchio}, {Avelino}, {Baeke}, {Bailli{\'e}}, {Balado}, {Ballber Balaguer{\'o}}, {Balestra}, {Ball}, {Ballans},
  {Ballot}, {Barban}, {Barbary}, {Barbieri}, {Barcel{\'o} Forteza}, {Barker}, {Barklem}, {Barnes}, {Barrado Navascues}, {Barragan}, {Baruteau}, {Basu}, {Baudin}, {Baumeister}, {Bayliss}, {Bazot}, {Beck}, {Belkacem}, {Bellinger}, {Benatti}, {Benomar}, {B{\'e}rard}, {Bergemann}, {Bergomi}, {Bernardo}, {Biazzo}, {Bignamini}, {Bigot}, {Billot}, {Binet}, {Biondi}, {Biondi}, {Birch}, {Bitsch}, {Bluhm Ceballos}, {B{\'o}di}, {Bogn{\'a}r}, {Boisse}, {Bolmont}, {Bonanno}, {Bonavita}, {Bonfanti}, {Bonfils}, {Bonito}, {Bonomo}, {B{\"o}rner}, {Boro Saikia}, {Borreguero Mart{\'\i}n}, {Borsa}, {Borsato}, {Bossini}, {Bouchy}, {Bou{\'e}}, {Boufleur}, {Boumier}, {Bourrier}, {Bowman}, {Bozzo}, {Bradley}, {Bray}, {Bressan}, {Breton}, {Brienza}, {Brito}, {Brogi}, {Brown}, {Brown}, {Brun}, {Bruno}, {Bruns}, {Buchhave}, {Bugnet}, {Buldgen}, {Burgess}, {Busatta}, {Busso}, {Buzasi}, {Caballero}, {Cabral}, {Cabrero Gomez}, {Calderone}, {Cameron}, {Cameron}, {Campante}, {Campos Gestal}, {Canto Martins}, {Cara}, {Carone}, {Carrasco},
  {Casagrande}, {Casewell}, {Cassisi}, {Castellani}, {Castro}, {Catala}, {Catal{\'a}n Fern{\'a}ndez}, {Catelan}, {Cegla}, {Cerruti}, {Cessa}, {Chadid}, {Chaplin}, {Charpinet}, {Chiappini}, {Chiarucci}, {Chiavassa}, {Chinellato}, {Chirulli}, {Christensen-Dalsgaard}, {Church}, {Claret}, {Clarke}, {Claudi}, {Clermont}, {Coelho}, {Coelho}, {Cogato}, {Colom{\'e}}, {Condamin}, {Conde Garc{\'\i}a}, \& {Conseil}}]{rau25}
{Rauer}, H., {Aerts}, C., {Cabrera}, J., {et~al.} 2025, Experimental Astronomy, 59, 26

\bibitem[{{Raymond} \& {Morbidelli}(2022)}]{ray22}
{Raymond}, S.~N. \& {Morbidelli}, A. 2022, in Astrophysics and Space Science Library, Vol. 466, Demographics of Exoplanetary Systems, Lecture Notes of the 3rd Advanced School on Exoplanetary Science, ed. K.~{Biazzo}, V.~{Bozza}, L.~{Mancini}, \& A.~{Sozzetti}, 3--82

\bibitem[{{Rice} {et~al.}(2022){Rice}, {Wang}, {Wang}, {Stef{\'a}nsson}, {Isaacson}, {Howard}, {Logsdon}, {Schweiker}, {Dai}, {Brinkman}, {Giacalone}, \& {Holcomb}}]{rice22wj}
{Rice}, M., {Wang}, S., {Wang}, X.-Y., {et~al.} 2022, \aj, 164, 104

\bibitem[{{Ricker} {et~al.}(2015){Ricker}, {Winn}, {Vanderspek}, \& et~al.}]{ricker15}
{Ricker}, G.~R., {Winn}, J.~N., {Vanderspek}, R., \& et~al. 2015, Journal of Astronomical Telescopes, Instruments, and Systems, 1, 014003

\bibitem[{{Romanova} {et~al.}(2013){Romanova}, {Ustyugova}, {Koldoba}, \& {Lovelace}}]{ro13}
{Romanova}, M.~M., {Ustyugova}, G.~V., {Koldoba}, A.~V., \& {Lovelace}, R.~V.~E. 2013, \mnras, 430, 699

\bibitem[{{Rossiter}(1924)}]{ross24}
{Rossiter}, R.~A. 1924, \apj, 60, 15

\bibitem[{{Sedaghati} {et~al.}(2023){Sedaghati}, {Jord{\'a}n}, {Brahm}, \& et~al.}]{seda23}
{Sedaghati}, E., {Jord{\'a}n}, A., {Brahm}, R., \& et~al. 2023, \aj, 166, 130

\bibitem[{{Shan} {et~al.}(2023){Shan}, {Yang}, {Lu}, {Wei}, {Tian}, {Zhang}, {Guo}, {Cui}, {Yang}, {Zhang}, \& {Liu}}]{shan21}
{Shan}, S.-S., {Yang}, F., {Lu}, Y.-J., {et~al.} 2023, \apjs, 264, 37

\bibitem[{{Siegel} {et~al.}(2023){Siegel}, {Winn}, \& {Albrecht}}]{sie23}
{Siegel}, J.~C., {Winn}, J.~N., \& {Albrecht}, S.~H. 2023, \apjl, 950, L2

\bibitem[{{Skarka} {et~al.}(2022){Skarka}, {{\v{Z}}{\'a}k}, {Fedurco}, {Paunzen}, {Henzl}, {Ma{\v{s}}ek}, {Karjalainen}, {Sanchez Arias}, {S{\'o}dor}, {Auer}, {Kab{\'a}th}, {Karjalainen}, {Li{\v{s}}ka}, \& {{\v{S}}tegner}}]{ska22}
{Skarka}, M., {{\v{Z}}{\'a}k}, J., {Fedurco}, M., {et~al.} 2022, \aap, 666, A142

\bibitem[{{Smith} {et~al.}(2019){Smith}, {Csizmadia}, {Gandolfi}, {Albrecht}, {Alonso}, {Barrag{\'a}n}, {Cabrera}, {Cochran}, {Dai}, {Deeg}, {Eigm{\"u}ller}, {Endl}, {Erikson}, {Fridlund}, {Fukui}, {Grziwa}, {Guenther}, {Hatzes}, {Hidalgo}, {Hirano}, {Korth}, {Kuzuhara}, {Livingston}, {Narita}, {Nespral}, {Niraula}, {Nowak}, {Palle}, {P{\"a}tzold}, {Persson}, {Prieto-Arranz}, {Rauer}, {Redfield}, {Ribas}, \& {Van Eylen}}]{smi19}
{Smith}, A.~M.~S., {Csizmadia}, S., {Gandolfi}, D., {et~al.} 2019, \actaa, 69, 135

\bibitem[{{Soto} {et~al.}(2018){Soto}, {D{\'\i}az}, {Jenkins}, {Rojas}, {Espinoza}, {Brahm}, {Drass}, {Jones}, {Rabus}, {Hartman}, {Sarkis}, {Jord{\'a}n}, {Lachaume}, {Pantoja}, {Vu{\v{c}}kovi{\'c}}, {Ciardi}, {Crossfield}, {Dressing}, {Gonzales}, \& {Hirsch}}]{soto18}
{Soto}, M.~G., {D{\'\i}az}, M.~R., {Jenkins}, J.~S., {et~al.} 2018, \mnras, 478, 5356

\bibitem[{{Southworth}(2011)}]{south11}
{Southworth}, J. 2011, \mnras, 417, 2166

\bibitem[{{Spalding} \& {Winn}(2022)}]{spal22}
{Spalding}, C. \& {Winn}, J.~N. 2022, \apj, 927, 22

\bibitem[{{Thies} {et~al.}(2011){Thies}, {Kroupa}, {Goodwin}, {Stamatellos}, \& {Whitworth}}]{th11}
{Thies}, I., {Kroupa}, P., {Goodwin}, S.~P., {Stamatellos}, D., \& {Whitworth}, A.~P. 2011, \mnras, 417, 1817

\bibitem[{{Thorngren}(2024)}]{tho24}
{Thorngren}, D.~P. 2024, arXiv e-prints, arXiv:2405.05307

\bibitem[{{Thygesen} {et~al.}(2023){Thygesen}, {Ranshaw}, {Rodriguez}, {Vanderburg}, {Quinn}, {Eastman}, {Bieryla}, {Latham}, {Vanderspek}, {Jenkins}, {Caldwell}, {Ikwut-Ukwa}, {Col{\'o}n}, {Dotson}, {Hedges}, {Collins}, {Calkins}, {Berlind}, \& {Esquerdo}}]{thy23}
{Thygesen}, E., {Ranshaw}, J.~A., {Rodriguez}, J.~E., {et~al.} 2023, \aj, 165, 155

\bibitem[{{Triaud}(2018)}]{tri17}
{Triaud}, A. H.~M.~J. 2018, in Handbook of Exoplanets, ed. H.~J. {Deeg} \& J.~A. {Belmonte}, 2

\bibitem[{{Turrini} {et~al.}(2021){Turrini}, {Schisano}, {Fonte}, \& et~al.}]{tur21}
{Turrini}, D., {Schisano}, E., {Fonte}, S., \& et~al. 2021, \apj, 909, 40

\bibitem[{{Winn} {et~al.}(2010){Winn}, {Fabrycky}, {Albrecht}, \& {Johnson}}]{winn10}
{Winn}, J.~N., {Fabrycky}, D., {Albrecht}, S., \& {Johnson}, J.~A. 2010, \apjl, 718, L145

\bibitem[{{Yang} {et~al.}(2024){Yang}, {Long}, {Kerins}, {Awiphan}, {Shan}, {Zhang}, {Joshi}, {A-thano}, {Jiang}, {Priyadarshi}, \& {Liu}}]{yan24}
{Yang}, F., {Long}, R.~J., {Kerins}, E., {et~al.} 2024, \mnras, 535, L7

\bibitem[{{Zak} {et~al.}(2024{\natexlab{a}}){Zak}, {Bocchieri}, {Sedaghati}, \& et~al.}]{zak24a}
{Zak}, J., {Bocchieri}, A., {Sedaghati}, E., \& et~al. 2024{\natexlab{a}}, \aap, 686, A147

\bibitem[{{Zak} {et~al.}(2025{\natexlab{a}}){Zak}, {Boffin}, {Bocchieri}, {Sedaghati}, {Balkoova}, \& {Kabath}}]{zak25b}
{Zak}, J., {Boffin}, H.~M.~J., {Bocchieri}, A., {et~al.} 2025{\natexlab{a}}, arXiv e-prints, arXiv:2505.20516

\bibitem[{{Zak} {et~al.}(2025{\natexlab{b}}){Zak}, {Boffin}, {Sedaghati}, {Bocchieri}, {Balkoova}, {Skarka}, \& {Kabath}}]{zak25a}
{Zak}, J., {Boffin}, H.~M.~J., {Sedaghati}, E., {et~al.} 2025{\natexlab{b}}, \aap, 694, A91

\bibitem[{{Zak} {et~al.}(2024{\natexlab{b}}){Zak}, {Boffin}, {Sedaghati}, \& et~al.}]{zak24b}
{Zak}, J., {Boffin}, H.~M.~J., {Sedaghati}, E., \& et~al. 2024{\natexlab{b}}, \aap, 687, L2

\bibitem[{{Zhou} {et~al.}(2018){Zhou}, {Rodriguez}, {Vanderburg}, {Quinn}, {Irwin}, {Huang}, {Latham}, {Bieryla}, {Esquerdo}, {Berlind}, \& {Calkins}}]{zhou18}
{Zhou}, G., {Rodriguez}, J.~E., {Vanderburg}, A., {et~al.} 2018, \aj, 156, 93

\end{thebibliography}

\begin{appendix}

\section{Additional tables and figures}
\label{mcmcsec}
\setcounter{figure}{0}
\renewcommand{\thefigure}{A\arabic{figure}}

\begin{figure}[h!]
\includegraphics[width=0.45\textwidth]{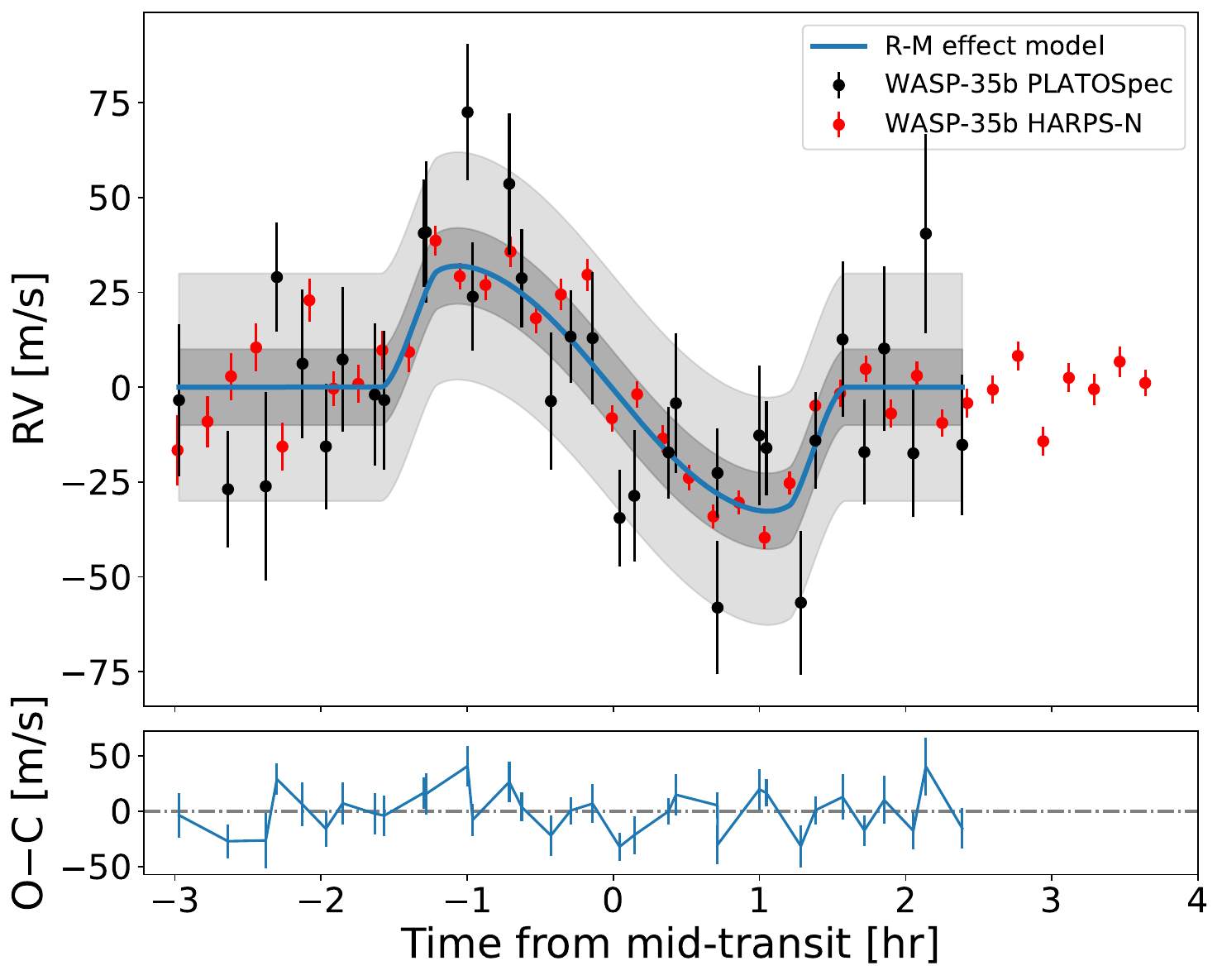}
\caption{Same as Fig. \ref{f:35PS} but with overlapping 3.6 m/HARPS-N data \citep{zak25b}. This illustrates the performance of PLATOSpec and the ability to obtain comparable results on 1.5m telescope with scheduling flexibility.}
\label{f:35PSH}
\end{figure}

\begin{figure}[h!]
\includegraphics[width=0.45\textwidth]{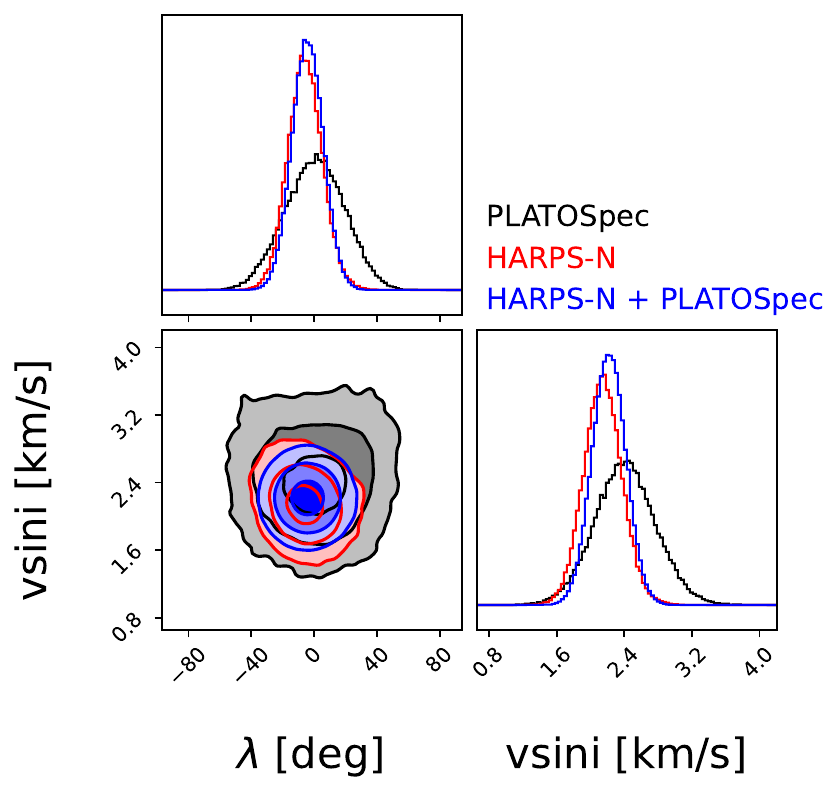}
\caption{Corner plot for WASP-35b showing posteriors for two parameters, projected spin-orbit angle $\lambda$ and projected rotational velocity of the star $v\,\sin{i_*}$. The contours show 1, 2 and 3-$\sigma$ deviations from the median. Data in red were obtained during a single transit with HARPS-N and presented in \citet{zak25b} while data in black are from two transits with the new PLATOSpec instrument. They show excellent agreement between the derived parameters. Data in blue are a joint fit of both datasets.}
\label{f:35PSHcomp}
\end{figure}

\begin{figure*}
\includegraphics[width=1\textwidth]{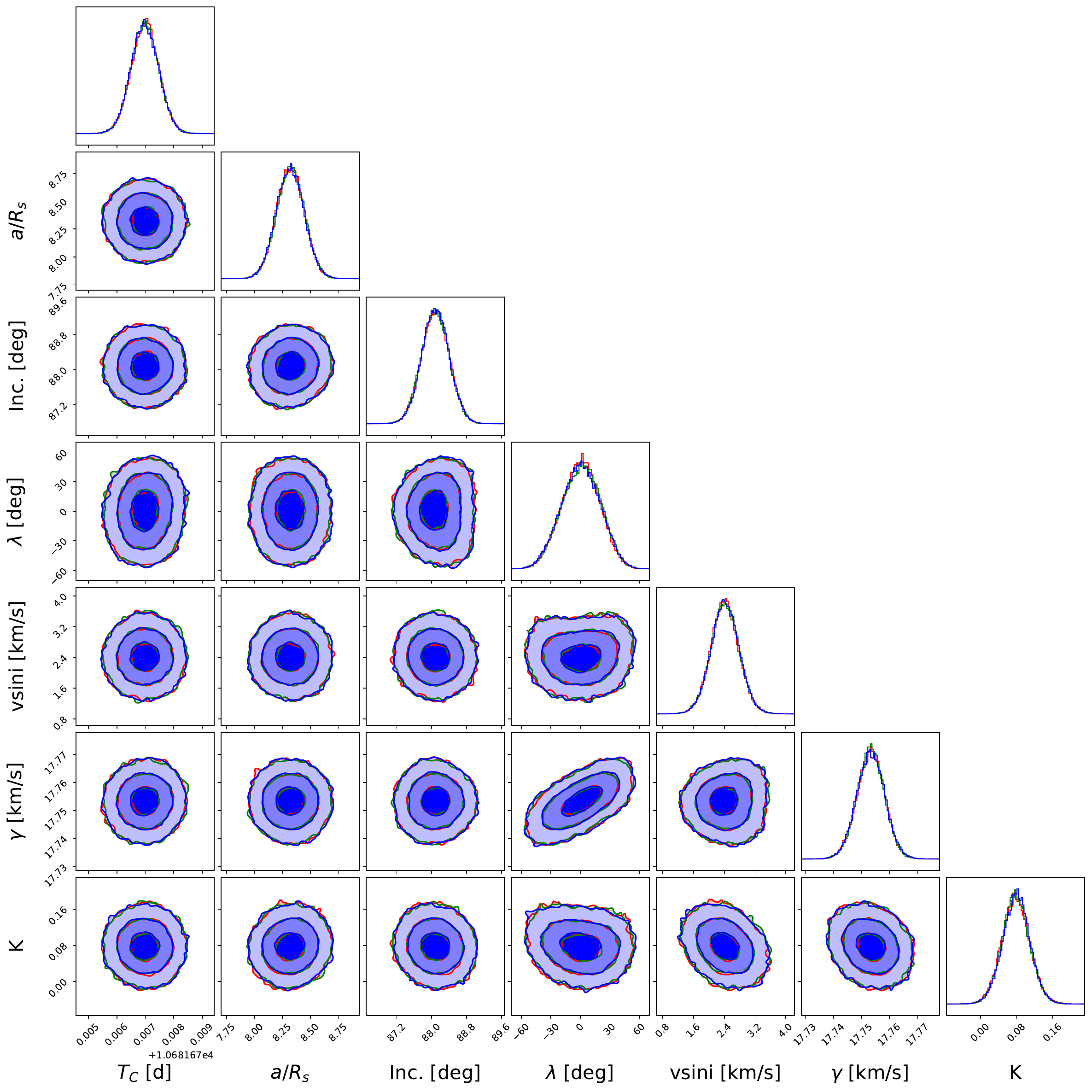}
\caption{MCMC results of WASP-35b. Three independent MCMC simulations are shown with different colors. The contours show 1, 2 and 3-$\sigma$ deviations from the median.}
\label{f:mcmc35}
\end{figure*}

\begin{figure*}
\includegraphics[width=1\textwidth]{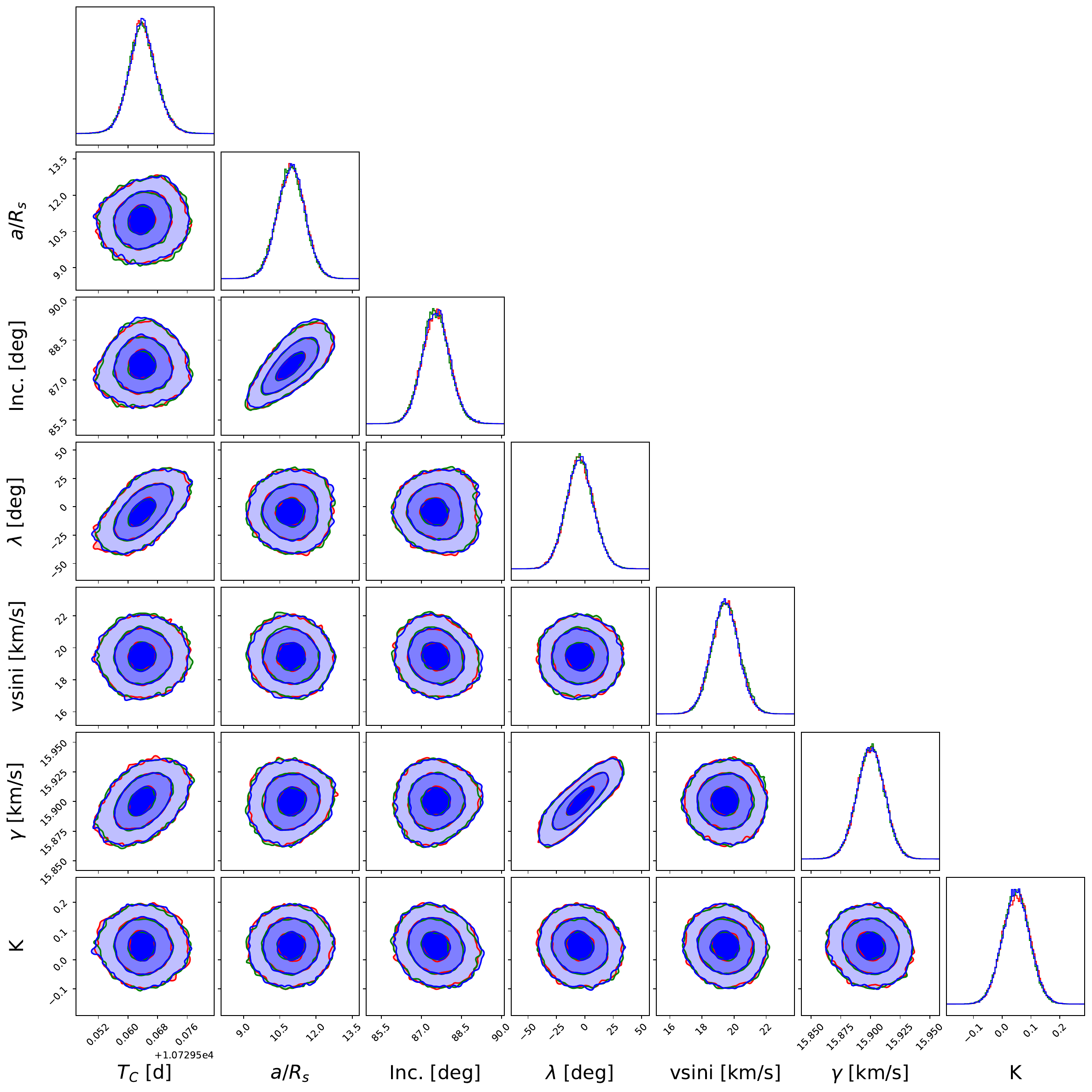}
\caption{MCMC results of TOI-622b. Three independent MCMC simulations are shown with different colors. The contours show 1, 2 and 3-$\sigma$ deviations from the median.}
\label{f:mcmc622}
\end{figure*}

\begin{figure*}
\includegraphics[width=1\textwidth]{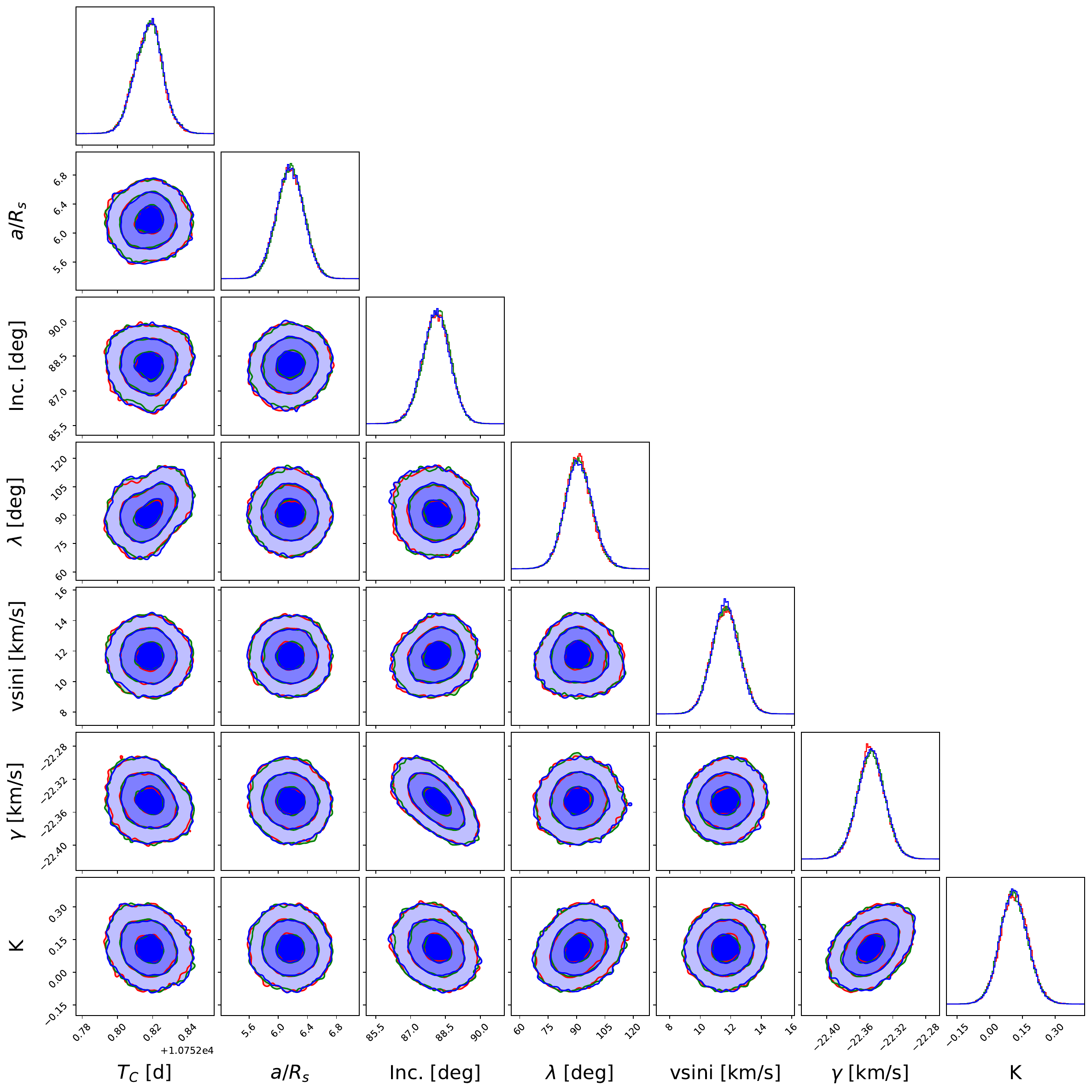}
\caption{MCMC results of K2-237b. Three independent MCMC simulations are shown with different colors. The contours show 1, 2 and 3-$\sigma$ deviations from the median.}
\label{f:mcmck2237}
\end{figure*}

\section{Derived radial velocities}
\label{appB}

\begin{table}[h!]
\caption{Radial velocity measurements for WASP-35.}
\label{table:rv35}
\centering
\begin{tabular}{c c c c}
\hline\hline
Date       & RV                     & RV uncertainty                   & Exp. time \\
$[\mathrm{BJD_{TDB}}]$    & $[\mathrm{km\,s^{-1}}]$ & $[\mathrm{km\,s^{-1}}]$ & $[\mathrm{s}]$     \\
\hline
2460681.55322686 &  17.7693 &  0.0201 &  1200 \\
2460681.56718455 &  17.7438 &  0.0154 &  1200 \\
2460681.58113067 &  17.7973 &  0.0143 &  1200 \\
2460681.59508836 &  17.7496 &  0.0166 &  1200 \\
2460681.60903448 &  17.7610 &  0.0187 &  1200 \\
2460681.62299217 &  17.8007 &  0.0142 &  1200 \\
2460681.63694986 &  17.7816 &  0.0143 &  1200 \\
2460681.65089597 &  17.7837 &  0.0129 &  1200 \\
2460681.66485365 &  17.7661 &  0.0123 &  1200 \\
2460681.67881133 &  17.7161 &  0.0128 &  1200 \\
2460681.69275744 &  17.7310 &  0.0121 &  1200 \\
2460681.70671511 &  17.7228 &  0.0118 &  1200 \\
2460681.72066122 &  17.7271 &  0.0124 &  1200 \\
2460681.73461889 &  17.7260 &  0.0128 &  1200 \\
2460681.74857656 &  17.7206 &  0.0138 &  1200 \\
2460681.76252266 &  17.7189 &  0.0168 &  1200 \\
2460681.77648033 &  17.7192 &  0.0185 &  1200 \\
2460700.54737477 &  17.7445 &  0.0249 &   900 \\
2460700.55784860 &  17.7750 &  0.0196 &   900 \\
2460700.56927143 &  17.7741 &  0.0191 &  1020 \\
2460700.58114562 &  17.7613 &  0.0183 &  1020 \\
2460700.59301981 &  17.8035 &  0.0186 &  1020 \\
2460700.60489400 &  17.8330 &  0.0180 &  1020 \\
2460700.61677975 &  17.8120 &  0.0186 &  1020 \\
2460700.62865394 &  17.7526 &  0.0181 &  1020 \\
2460700.64052812 &  17.7671 &  0.0174 &  1020 \\
2460700.65240231 &  17.7234 &  0.0173 &  1020 \\
2460700.66427649 &  17.7457 &  0.0185 &  1020 \\
2460700.67615067 &  17.6897 &  0.0176 &  1020 \\
2460700.68802484 &  17.7330 &  0.0185 &  1020 \\
2460700.69989902 &  17.6868 &  0.0190 &  1020 \\
2460700.71177320 &  17.7541 &  0.0205 &  1020 \\
2460700.72364738 &  17.7496 &  0.0216 &  1020 \\
2460700.73552155 &  17.7778 &  0.0262 &  1020 \\
\hline
\end{tabular}
\end{table}

\begin{table}[h!]
\caption{Radial velocity measurements for TOI-622.}
\label{table:rv622}
\centering
\begin{tabular}{c c c c}
\hline\hline
Date       & RV                     & RV uncertainty                    & Exp. time \\
$[\mathrm{BJD_{TDB}}]$    & $[\mathrm{km\,s^{-1}}]$ & $[\mathrm{km\,s^{-1}}]$ & $[\mathrm{s}]$     \\
\hline
2460716.76909215 &  15.8986 &  0.0316 &  900 \\
2460716.77957833 &  15.8566 &  0.0304 &  900 \\
2460716.79005294 &  15.8478 &  0.0284 &  900 \\
2460716.80053913 &  15.8044 &  0.0298 &  900 \\
2460716.81102531 &  15.8440 &  0.0314 &  900 \\
2460716.82149992 &  15.8159 &  0.0317 &  900 \\
2460716.83198610 &  15.8111 &  0.0325 &  900 \\
2460716.84247228 &  15.8752 &  0.0305 &  900 \\
2460716.85338670 &  15.8917 &  0.0299 &  900 \\
2460716.86386131 &  15.8106 &  0.0317 &  900 \\
2460716.87434749 &  15.7563 &  0.0320 &  900 \\
2460793.48528093 &  15.9171 &  0.0316 &  900 \\
2460793.49285585 &  15.8756 &  0.0331 &  900 \\
2460793.49990417 &  15.9178 &  0.0299 &  900 \\
2460793.50877532 &  15.9482 &  0.0245 &  900 \\
2460793.51926099 &  15.9609 &  0.0246 &  900 \\
2460793.52974667 &  16.0018 &  0.0239 &  900 \\
2460793.54022076 &  16.0276 &  0.0234 &  900 \\
2460793.55070643 &  15.9978 &  0.0239 &  900 \\
2460793.56119210 &  15.9916 &  0.0263 &  900 \\
2460793.57167777 &  15.9435 &  0.0252 &  900 \\
2460793.58215187 &  15.9236 &  0.0242 &  900 \\
2460793.59263754 &  15.8276 &  0.0241 &  900 \\
2460793.60312320 &  15.8782 &  0.0242 &  900 \\
2460793.61359730 &  15.8979 &  0.0243 &  900 \\
2460793.62408296 &  15.8046 &  0.0243 &  900 \\
2460793.63456863 &  15.8542 &  0.0252 &  900 \\
2460793.64504272 &  15.8123 &  0.0258 &  900 \\

\hline
\end{tabular}
\end{table}

\begin{table}[h!]
\caption{Radial velocity measurements for K2-237.}
\label{table:rvk2237}
\centering
\begin{tabular}{c c c c}
\hline\hline
Date       & RV                     & RV uncertainty                    & Exp. time \\
$[\mathrm{BJD_{TDB}}]$    & $[\mathrm{km\,s^{-1}}]$ & $[\mathrm{km\,s^{-1}}]$ & $[\mathrm{s}]$     \\
\hline
2460752.74151981 & -22.2392 & 0.0670 & 1200 \\
2460752.75546794 & -22.3986 & 0.0487 & 1200 \\
2460752.77364103 & -22.3856 & 0.0333 & 1750 \\
2460752.79395552 & -22.4809 & 0.0291 & 1750 \\
2460752.81428159 & -22.3529 & 0.0337 & 1750 \\
2460752.83459609 & -22.4147 & 0.0326 & 1750 \\
2460752.85491058 & -22.5531 & 0.0268 & 1750 \\
2460752.87523664 & -22.3925 & 0.0266 & 1750 \\
2460752.89309718 & -22.4353 & 0.0333 & 1300 \\
2460752.90821439 & -22.4511 & 0.0369 & 1300 \\
2460787.65942431 & -22.4138 & 0.0456 & 1600 \\
2460787.67801357 & -22.5013 & 0.0408 & 1600 \\
2460787.69660284 & -22.3185 & 0.0400 & 1600 \\
2460787.71519210 & -22.3788 & 0.0386 & 1600 \\
2460787.73378136 & -22.4204 & 0.0420 & 1600 \\
2460787.75247479 & -22.4254 & 0.0409 & 1600 \\
2460787.77124925 & -22.3107 & 0.0462 & 1600 \\
2460787.79009315 & -22.3039 & 0.0451 & 1600 \\
2460787.80867082 & -22.2765 & 0.0447 & 1600 \\
2460787.82726007 & -22.3319 & 0.0474 & 1600 \\
2460787.84590719 & -22.3789 & 0.0441 & 1600 \\
2460787.86539927 & -22.3262 & 0.0475 & 1600 \\
2460800.68198632 & -22.3043 & 0.0512 & 1200 \\
2460800.69594538 & -22.3384 & 0.0513 & 1200 \\
2460800.70989287 & -22.2357 & 0.0548 & 1200 \\
2460800.72385194 & -22.3942 & 0.0534 & 1200 \\
2460800.73779942 & -22.3015 & 0.0516 & 1200 \\
2460800.75175848 & -22.3724 & 0.0555 & 1200 \\
2460800.76570597 & -22.4482 & 0.0543 & 1200 \\
2460800.77966502 & -22.5075 & 0.0550 & 1200 \\
2460800.79362408 & -22.3671 & 0.0559 & 1200 \\
2460800.80757156 & -22.4988 & 0.0574 & 1200 \\
2460800.82153061 & -22.3526 & 0.0536 & 1200 \\
2460800.83547808 & -22.4352 & 0.0674 & 1200 \\
2460800.84943713 & -22.3298 & 0.0527 & 1200 \\
2460800.86338460 & -22.3894 & 0.0583 & 1200 \\
2460800.87734365 & -22.3674 & 0.0577 & 1200 \\
2460800.89129112 & -22.5943 & 0.0639 & 1200 \\
2460800.90525016 & -22.4263 & 0.0620 & 1200 \\
2460800.91919762 & -22.3268 & 0.0716 & 1200 \\
\hline
\end{tabular}
\end{table}

\end{appendix}
\end{document}